\documentclass[journal,10pt]{IEEEtran}

\usepackage{xcolor,soul,framed} %,caption
\colorlet{shadecolor}{yellow}
\usepackage[pdftex]{graphicx}
\graphicspath{{../pdf/}{../jpeg/}}
\DeclareGraphicsExtensions{.pdf,.jpeg,.png .jpg}
\usepackage{float}
\usepackage{hyperref}                     % Loaded by Thesis.cls
\usepackage{subcaption}
\usepackage{algpseudocode}
\usepackage[cmex10]{amsmath}
\usepackage{flexisym}
\usepackage{array}
\usepackage{mdwmath}
\usepackage{mdwtab}
\usepackage{xcolor}

\newsavebox\MBox
\usepackage[linesnumbered,ruled,vlined]{algorithm2e}
\SetCommentSty{mycommfont}
\SetKwInput{KwInput}{Input}                % Set the Input
\SetKwInput{KwOutput}{Output}              % set the Output
\SetKwFunction{KwAgentUpdate}{AgeUpdate}
\SetKwFunction{FMain}{AgentUpdate}
\SetKwProg{Fn}{Function}{:}{}
\usepackage{algpseudocode}
\makeatletter
\def\BState{\State \hskip-\ALG@thistlm}
\makeatother

\usepackage{algpseudocode}
\usepackage{eqparbox}
\usepackage{url}
\usepackage[cmex10]{amsmath}
\usepackage{cite}
\usepackage{mathtools}
\usepackage{graphicx}
\usepackage{amsmath}
\usepackage{amssymb}
\usepackage{pifont}% http://ctan.org/pkg/pifont
\usepackage{caption}
\IEEEoverridecommandlockouts

\usepackage{atbegshi}% To remove first blank page 
\usepackage{adjustbox}

\begin{document}

\title{AI-based Radio and Computing Resource Allocation and Path Planning in NOMA NTNs:\\AoI Minimization under CSI Uncertainty}

\author{Maryam~Ansarifard,
	Nader~Mokari,~\IEEEmembership{Senior Member,~IEEE,}
	Mohammadreza~Javan,~\IEEEmembership{Senior Member,~IEEE,}
	Hamid~Saeedi,~\IEEEmembership{Senior Member,~IEEE,}
	and~Eduard~A. Jorswieck,~\IEEEmembership{Fellow,~IEEE}
	% <-this % stops a space
	\thanks{M. Ansarifard, N. Mokari and H. Saeedi are with the Department of Electrical and Computer Engineering, Tarbiat Modares University, Tehran, Iran, (emails: {m.ansarifard, nader.mokari, hsaeedi}@modares.ac.ir). H. Saeedi is also with University of Doha for Science and Technology, Doha, Qatar. M. R. Javan is with the Department of Electrical Engineering, Shahrood University of Technology, Iran, (email: javan@shahroodut.ac.ir). Eduard A. Jorswieck is
		with Institute for Communications Technology, TU Braunschweig, Germany,
		email: jorswieck@ifn.ing.tu-bs.de.}}

\markboth{IEEE Transactions on Vehicular Technology}%,~Vol.~XX, No.~XX, XXX~2015}
{}
%{Shell \MakeLowercase{\textit{et al.}}: Bare Demo of IEEEtran.cls for Journals}

\maketitle

\begin{abstract}
%Non-terrestrial networks (NTNs) are emerging as a promising solution to handle the increasing demand of computation requirements. Due to the importance of latency in computation-intensive applications, the criterion of age of information (AoI) has been introduced, which gives a better view of the freshness of information. 
In this paper, we develop a hierarchical aerial computing framework composed of high altitude platform (HAP) and unmanned aerial vehicles (UAVs) to compute the fully offloaded tasks of terrestrial mobile users which are connected through an uplink non-orthogonal multiple access (UL-NOMA). To better assess the freshness of information in computation-intensive applications the criterion of age of information (AoI) is considered. In particular, the problem is formulated to minimize the average AoI of users with elastic tasks, by adjusting UAVs' trajectory and resource allocation on both UAVs and HAP, which is restricted by the channel state information (CSI) uncertainty and multiple resource constraints of UAVs and HAP. In order to solve this non-convex optimization problem, two methods of multi-agent deep deterministic policy gradient (MADDPG) and federated reinforcement learning (FRL) are proposed to design the UAVs' trajectory, and obtain channel, power, and CPU allocations. It is shown that task scheduling significantly reduces the average AoI. This improvement is more pronounced for larger task sizes. On one hand, it is shown that power allocation has a marginal effect  on the average AoI compared to using full transmission power for all users. Compared with traditional transmission schemes, the simulation results show our scheduling scheme results in a substantial improvement in average AoI.
\end{abstract}


\IEEEpeerreviewmaketitle

\section{Introduction}

\subsection{State-of-the-Art}
In the next generation of wireless networks, developing various traffic types and use cases requires a greater degree of planning and flexibility. As a result of their strong growth in the coming years, internet of things environments (IoT) are the top priorities of these requirements, ranging from machine-type communications to ultra-reliable and low-latency communications, industrial scenarios, and novel applications requiring high throughput. A new cellular IoT technology characterizing reduced-capability (RedCap) user equipments for the fifth-generation (5G) new radio (NR) is introduced in the 3rd generation partnership project (3GPP) Rel-17, which specifies the device capabilities required to support novel mid-end IoT use cases, including urban monitoring or video surveillance, wearable and industrial wireless sensors \cite{3gpp.38.875, whitepaper}. As several services %defined in [7] (i.e., process management, work safety, and healthcare monitoring)
require real-time uplink (UL) data transmission, meeting the stringent requirements of the above industrial applications prompts new solutions in terms of offloading Redcap devices' computation-intensive tasks. Non-terrestrial networks (NTNs), with using elements such as unmanned aerial vehicles (UAVs), high altitude platforms (HAPs), and satellites provide a 3D flexible coverage. Moreover, as the devices on the ground are unable to perform computationally-expensive operations, HAPs and UAVs, equipped with computation resources, are considered as potential candidates for mobile edge computing (MEC) \cite{kurt2021vision}. A fixed HAP remains at an altitude of around $20$$\text{km}$ equipped with powerful loading equipment such as computing devices, which provides huge coverage, making it an ideal base station in the air. However, terrestrial devices with limited power supplies cannot be connected directly to HAPs. Such a connection can be made through UAVs.
Due to their autonomy, flexibility, and wide range of application domains, UAVs have gained popularity over the past few years \cite{introUAV1,introUAV2}. To enhance the coverage and communication quality, UAVs are employed as mobile base stations \cite{introuavbs1, introuavbs2}. Also, in order to reduce the access latency and to improve the communication quality of the UAV networks, non-orthogonal multiple access (NOMA) can be applied to the UAV networks. \\%In case of tasks with heavy computational and storage requirements, UAVs with limited resources cannot complete the whole task. Therefore, it is necessary for UAVs to cooperate with HAPs. \\%; as HAPs can provide extra computing capacity. \\ 
\indent The traditional approach of offloading tasks to the NTN, whereby the whole task would be sent at once as long as sufficient resources are available, can result in large transmission latency that degrades users' quality of experience (QoE). Therefore, a time-aware task scheduling considering available radio and computation resources is needed. In addition to  performing the computations of the received tasks, the freshness of the information is very crucial in delay-sensitive applications. To measure the performance of data freshness at the receiver side, a new metric, called age of information (AoI), is proposed in \cite{introAoI1}, which is defined as the amount of time elapsed since the freshest delivered update takes place. Many recent works have focused on minimizing this metric in different networks \cite{introAoI2,introAoI3}. However, there are some factors that affect this metric, in particular, imperfect channel state information (CSI) resulting from estimation errors and limited feedback from users in practice \cite{CSI1}. Consequently, it is pertinent to take into account the imperfection of CSI in links between ground users and UAVs.\\
\indent In uncertain environments where the CSI is not perfectly known and for time-sensitive applications, managing the resources and mobility of non-terrestrial platforms is a challenging task. Conventional mathematical optimization approaches may fail to converge within the desired time range for these problems, which are usually non-convex \cite{naous2023reinforcement}. Artificial intelligence (AI) and in particular reinforcement learning (RL) algorithms have been developed in NTN-aided wireless communications and cellular-connected NTNs over the last few years, where by continuously interacting with the environment, the agent will learn a certain skill by using the reward value provided by the environment. The concept of RL has been extensively studied and applied, and it is considered to be one of the core technologies of intelligent systems design. Due to federated learning (FL) success in supervised learning tasks, federated RL (FRL) has become an attractive subject. Using FL, raw data generated locally can be used to train a local model, and then the local model parameters can be used to update the weights of the global model or other local models (in peer-to-peer approach). The multi-agent deep deterministic policy gradient (MADDPG) and FRL can be applied to cooperative, competitive, or mixed multi-agent environments \cite{introMADDPG}. Although the agent can make decisions and act independently, the environment and other agents will impact its decisions and actions.\\
\indent This article aims to minimize the average AoI of all ground users in proposed hierarchical NTN architecture, including multi UAVs and a HAP equipped with computational resources to process the received tasks, by elastic task scheduling, radio and computing resource allocation of UAVs and the HAP, and trajectory planning of UAVs. To examine the effect of CSI imperfection, we consider CSI uncertainty in uplink NOMA between ground users and UAVs, and in order to solve this non-convex problem, we adopt MADDPG and FRL approches.

\subsection{Related Works}
UAV-assisted systems have been combined with other technologies such as NOMA. In \cite{rnoma1, rnoma2, rnoma3} NOMA is employed to improve the performance of UAV-enabled communication systems. A swarm of UAVs is considered in \cite{rswarm} to ensure the long-term freshness requirements of situation awareness. To reduce energy consumption under fast-changing environmental dynamics, the authors use a multi-agent deep reinforcement learning (DRL) algorithm with global-local rewards. Based on UAV latency requirements and UAV location, \cite{yang2019energy} optimizes the total energy consumption of UAVs and users in a multiple UAV-enabled MEC networks planning. Ref. \cite{yang2020offloading} focuses on edge computing for UAVs to track a mobile target and identify it, satisfying stringent latency requirements. Overall cost and inference error have been traded off. On the other hand, HAPs can provide intensive computing services due to their stronger payload. Using a FL-based algorithm, \cite{wang2021federated} has designed a task computation algorithm for high-altitude computing-enabled balloons to minimize energy and time usage during the data offloading process. 
%A network composed of HAPs has been presented in \cite{ke2021edge}, aiming to provide massive IoT users with efficient connections and low latency. 
The hierarchical aerial computing framework developed by \cite{jia2022hierarchical}, provides MEC services for various IoT applications to maximize total IoT data computed, which is constrained by IoT delays and UAV and HAP resource requirements. In \cite{peng2022deep}, a deep Q-learning network (DDQN) algorithm is proposed to realize the freshness-aware path planning of the UAV. 
In IoT systems, AoI-aware UAV-aided wireless network has received significant attention. In \cite{raio2}, the UAV's trajectory and data acquisition mode are optimized in accordance with the aim of minimizing the average AoI of all sensors. Ref. \cite{raoi1} uses the UAV to collect the sensors' data, and the UAV's trajectory is optimized to minimize the average AoI of the system as well as the maximum AoI of the individual wireless sensors. In \cite{raoi3} a trajectory planning strategy for UAVs is proposed to minimize the maximum AoI of a UAV-enabled wireless sensor network, ensuring the well-balanced between the sensors' uploading time and the UAV’s flight time. A decentralized computation offloading algorithm is proposed in \cite{wang2020multi} with the aim of minimizing average task completion time. Ref. \cite{hu2020aoi} minimizes the average AoI of the data collected from all ground sensors by optimizing joint energy transfer and data collection time allocation and UAV’s trajectory planning, where the authors decomposed it into two sub-problems and solved it with designing dynamic programming (DP) and ant colony (AC)  heuristic algorithms. A distributed optimization problem for resource allocation at the MEC servers is formulated in  \cite{Peng2021MultiAgentRL} to maximize the number of offloaded tasks while satisfying heterogeneous quality-of-service (QoS) requirements, and then it is solved using the MADDPG. In \cite{zhu2021federated}, the authors propose a heterogeneous multi-agent actor-critic algorithm based on RL to minimize the average AoI in MEC systems.
%\IEEEpubidadjcol
%CSI_______________
The reality of UAV missions usually involves an uncertain and dynamic environment, so the control architecture must be robust and versatile enough to deal with uncertainty and changes in the environment \cite{run1}. Currently, AoI-aware problems have mainly focused on maximizing total data rates of users under perfect CSI. Authors in \cite{rcsi2} study the average staleness of CSI in fully connected time-varying reciprocal wireless networks with node equi-presumably selected to transmit and distribute CSI in each time slot. In Table \ref{tab:my-table}, we have summarized the related works with respect to our proposed framework. As can be seen, in the aforementioned literature, the concentration is on minimizing AoI through UAV trajectory and resource allocation by imposing transmission time constraints. Furthermore, they do not take into account the CSI uncertainty, which is inevitable in the proposed environment. Moreover, Redcap devices' servers have limited computing resources, and this limitation has rarely been paid attention to in these studies. 
	Overall, it appears that AoI-aware scheduling and resource allocation for resource-constrained users and NTN elements remains a challenge. %In the aforementioned literature, most concentration is only on minimizing AoI by UAV trajectory and resource allocation, imposing transmission time constraint. Moreover, they do not consider the impact of CSI uncertainty which is inevitable in the environment. However, these studies rarely take into account the limitation of computing resources of MEC servers. It seems that, the AoI-aware scheduling and resource allocation for the resource-constrained users and NTN elements remains an open problem so far. 
To the best of our knowledge, our work is the first one on minimizing average AoI up to the top layer, i.e., HAP, coupled with CSI imperfection on the user-UAV links, using joint trajectory planning, task scheduling, and resource allocation.
\begin{table*}[]
	\centering
	\caption{Related works comparison.}
	\label{tab:my-table}
	\scalebox{0.9}{
	\begin{tabular}{|c|c|c|c|c|c|c|c|c|}
		\hline
		Ref.     & NTN       & \begin{tabular}[c]{@{}c@{}}Resource \\ Allocation\end{tabular} & \begin{tabular}[c]{@{}c@{}}Task  \\ Offloading\end{tabular} & \begin{tabular}[c]{@{}c@{}}Task Transmission \\ Method\end{tabular}& \begin{tabular}[c]{@{}c@{}}Trajectory \\ Planning\end{tabular} & Uncertainty & Objective Function                                                                                        & Solution         \\ \hline 
		\cite{wang2021federated}\begin{tabular}[c]{@{}c@{}}\end{tabular}
		& \begin{tabular}[c]{@{}c@{}}High altitude \\ balloons (HABs)\end{tabular}                & $\checkmark$                                                               & $\checkmark$         & Fixed & \ding{55}                                                               & \ding{55}            & \begin{tabular}[c]{@{}c@{}}Minimizing weighted\\ sum of energy and \\ time consumption\end{tabular} & SVM-based FL     \\ \hline
		\cite{jia2022hierarchical}
		& UAVs and HAPs              & $\checkmark$                                                               & $\checkmark$              & Fixed & \ding{55}                                                              & \ding{55}           & \begin{tabular}[c]{@{}c@{}}Maximizing the total\\IoT computed data \end{tabular}                                                                                                    & \begin{tabular}[c]{@{}c@{}}Game theory,\\heuristic algorithm\end{tabular}           \\ \hline
		\cite{peng2022deep}
		& Single UAV               & $\checkmark$                                                               & $\checkmark$         & Fixed & $\checkmark$                                                               & \ding{55}            & \begin{tabular}[c]{@{}c@{}}Minimizing average AoI \\ and energy consumption\end{tabular} & DDQN    \\\hline
		\cite{raoi1}
		& Single UAV               & \ding{55}                                                             &$\checkmark$         & Fixed & $\checkmark$                                                               & \ding{55}            & \begin{tabular}[c]{@{}c@{}}Minimizing max/ \\ average AoI\end{tabular} & DP, genetic
		algorithm (GA)  \\ \hline 
		\cite{wang2020multi}
		& \ding{55}               & \ding{55}                                                               &$\checkmark$         & Fixed & \ding{55}                                                               & \ding{55}           & \begin{tabular}[c]{@{}c@{}}Minimizing average task \\ completion time\end{tabular} & MA imitation learning    \\ \hline \cite{hu2020aoi}
		& Single UAV               & $\checkmark$                                                               &$\checkmark$         & Fixed & $\checkmark$                                                               & \ding{55}           & Minimizing average AoI & DP, AC \\ \hline  
		\cite{Peng2021MultiAgentRL}
		& \begin{tabular}[c]{@{}c@{}}UAV-assisted \\ vehicular network\end{tabular}               & $\checkmark$                                                               & $\checkmark$         & Fixed & $\checkmark$                                                               & \ding{55}            & \begin{tabular}[c]{@{}c@{}}Maximizing the number \\ of offloaded tasks\end{tabular} & MADDPG     \\\hline  \cite{zhu2021federated}
		& Multi-UAVs        & $\checkmark$                                                        &$\checkmark$              & Fixed & $\checkmark$                                                               & \ding{55}          & Minimizing average AoI                                                                                                    & MADDPG, EdgeFed                  \\ \hline
		\begin{tabular}[c]{@{}c@{}}Our\\ work \end{tabular} & \begin{tabular}[c]{@{}c@{}}Both UAVs and \\ HAP\end{tabular} & $\checkmark$                                                              & $\checkmark$               & Elastic & $\checkmark$                                                        & $\checkmark$     & Minimizing average AoI                                                                                      & MADDPG, P2P-VFRL \\ \hline
	\end{tabular}
		}
\end{table*}

\subsection{Our Contribution}
%\textcolor{red}{Unlike the aforementioned literature, we establish a task scheduling in which there is no assumption on sending time of each user's task. This means that a part of the task or the whole task can be transferred in one time slot.}
%Motivated by the aforementioned literature, the main contribution of this paper is minimizing the AoI without knowing the whole sending time of each user's task. This means that a part of the task or the whole task can be transferred in one time slot and there is no constraint on transmission time. 
The key contributions of this work are as follows:
\begin{itemize}
	\item We first formulate the trajectory and resource allocation problem in a NOMA-enabled NTN. The aim of the formulated problem is to minimize the average AoI experienced by all the users under CSI uncertainty while satisfying the minimum data rate constraint of users. We notice that the formulated problem is non-convex, and mixed-integer nonlinear problem (MINLP) which cannot be solved globally and efficiently.
	%\item In this paper, we consider CSI uncertainty in uplink between users and UAVs, by constraining outage probability on experienced data rates.
	\item In the existing literature, receiving tasks are constrained by a time limit. This existing scheme is herein referred to as \emph{fixed method}. Due to the low latency and large data volume requirements of RedCap services, the \emph{fixed method} falls short in terms of delivering the whole task, resulting in an increased average AoI. To tackle this issue and to guarantee task timeliness, we propose a novel dynamic task transmission method that, unlike previous studies, does not assume the availability of the required radio and computing resources for the whole task to be offloaded. Our proposed method is compared against the \emph{fixed method} and it is shown that our method performs better in terms of minimizing average AoI.
	\item In the simulation, we deal with two types of agents, UAV and HAP. They differ in their actions but have the same goal, which is minimizing average AoI. Using online AI-based algorithms, we first present the convergence of the MADDPG approach. Then, we compare the average AoI experienced by the mobile users with a FRL-based algorithm, a peer-to-peer vertical federated reinforcement learning (P2P-VFRL). To the best of our knowledge, this is the first algorithm that combines the peer-to-peer vertical federated (P2P-VF) with the multi-agent actor-critic reinforcement learning, which is dynamically adapted to our problem.
	%		40.7 and 4.3 less than that of the greedy scheme and
	%		exhaustive search, respectively. Furthermore, the results
	%		also reveal that the average latency experienced by the
	%		users under our proposed algorithm is identical to that of
	%		the centralized scheme
\end{itemize}
\subsection{Paper Organization}
The rest of this paper is organized as follows. The system
model and the problem formulation are presented in Section \ref{II}. In Section \ref{IV}, the solution to the 
problem is provided. Simulation results are presented in Section \ref{V}, and Section \ref{VI} concludes the paper.\\
\textbf{Symbol Notations}: Column vectors and matrices are denoted by
boldface lower case letters and capital letters, respectively. The set of elements is denoted by calligraphic letters. The
transpose of a vector $\mathbf{a}$ is denoted by $\mathbf{a}^{T}$. The Euclidean norm of a vector and degenerate interval are denoted by $\|\cdot\|$ and $\left[a, b\right] = \left\{x | a \leq x \leq b\right\}$, respectively. 

\section{System Model} \label{II}
\subsection{UAV-BSs and Users}
We consider a HAP and multiple UAVs as aerial BSs, which are denoted by $\mathcal{U} = \{1, \dots, U\}$, where $U$ is the total number of UAV-BSs. We assume that we have $M$ users in our environment, $\mathcal{M}=\{1, \dots, M\}$ and each UAV-BS can serve at most $L$ users, and we denote $L_{u}$ to be the maximum number of users allocated to UAV-BS $u$.
%a set of $\mathcal{M}$, in which $\mathcal{M}=\{1, ..., M\}$ and it varies with time, where $M$ is the maximum number of users that each UAV-BS can have.
Time is considered to be slotted and normalized to the slot duration (i.e., the slot duration is taken as 1), and the UAV-BSs’ operating period is $T$ time slots, where $\mathcal{T}=\{0, 1, \dots, T\}$. The coordinates of the $m$-th user can be expressed as $\mathbf{v}_{m}(t) =[x_{m}(t),y_{m}(t)]^{T}$, where $x_{m}(t)$ and $y_{m}(t)$ are the X-coordinate and Y-coordinate of user $m$ in time slot $t$, respectively. The location of the HAP is denoted by $\mathbf{q}_{\text{HAP}} =[x_{\text{HAP}}, y_{\text{HAP}}, h_{\text{HAP}}]^{T}$, which is considered constant during the operation time of the network, and the location of UAV-BS $u$ is denoted by $\mathbf{q}_{u}(t) =[x_{u}(t), y_{u}(t), h_{u}(t)]^{T}$, in which $h_{u}(t) \in[h_{\text{min}},h_{\text{max}}]$. %Furthermore, we assume each UAV-BS should return to its initial location by the end of each flight time $T_{f}$. Hence, following constraint need to be considered \ref{return to initial location}
%\begin{equation}\label{return to initial location}
%	\mathbf{q}_{u}(0) = \mathbf{q}_{u}(T_{f}). \\
%\end{equation}
Furthermore, the trajectories of UAV-BSs are also subject to the collision avoidance constraint, i.e,
\begin{equation}\label{collision avoidance}
	C 1:\|\mathbf{q}_{u}(t)-\mathbf{q}_{u^{\prime}}(t)\|\geq d_{\text{min}} , \forall u\neq u^{\prime}, \forall t \in \mathcal{T}, \\
\end{equation}
where $d_{\text{min}}$ denotes the minimum inter UAV-BS distance. Let $v_{u}(t)$ be the velocity of UAV-BS $u$ at time slot $t$. As a result of the mechanical limitation, the maximum speed of each UAV-BS is $v_{\text{max}}$.
%and we assume this velocity is equal for all UAV-BSs i.e. $v_{u}=v_{u^{\prime}}, u \neq u^{\prime}$. The location of UAV-BS $u$ in time slot $t+1$ is given as \begin{equation}\mathbf{q}_{u}(t+1) = \mathbf{q}_{u}(t) + v_{u}(t). \omega_{u}(t)	\end{equation}where $\omega_{u}(t)$ is the trajectory direction of UAV-BS $u$ in time slot $t$.
%\begin{figure}[h]
%	\centering
%	\captionsetup{justification=centering}
%	\includegraphics[width=\linewidth]{pdf/Time Frame.pdf}
%	\caption{UAV-BSs' operation time.}
%	\label{timepic}
%\end{figure}
We assume that the initial location of UAV-BSs is random and users are already assigned to UAV-BSs based on the quality of their channels; and by that, we have a distance-based coverage area for each UAV-BS to serve its users. As a result, we have the following constraints:
\begin{equation}\label{clustring}
	C 2:\begin{cases}
		\min\limits_{m\in \mathcal{M}}{x_{m}}(t)\leq x_{u}(t)\leq \max\limits_{m\in \mathcal{M}}{x_{m}}(t),\\
		\min\limits_{m\in \mathcal{M}}{y_{m}}(t)\leq y_{u}(t)\leq \max\limits_{m\in \mathcal{M}}{y_{m}}(t).
	\end{cases}
\end{equation}
% [width=2\linewidth]
\begin{figure*}[t]
	\centering
	%\captionsetup{justification=centering}
	\includegraphics[width=\textwidth, height=10cm]{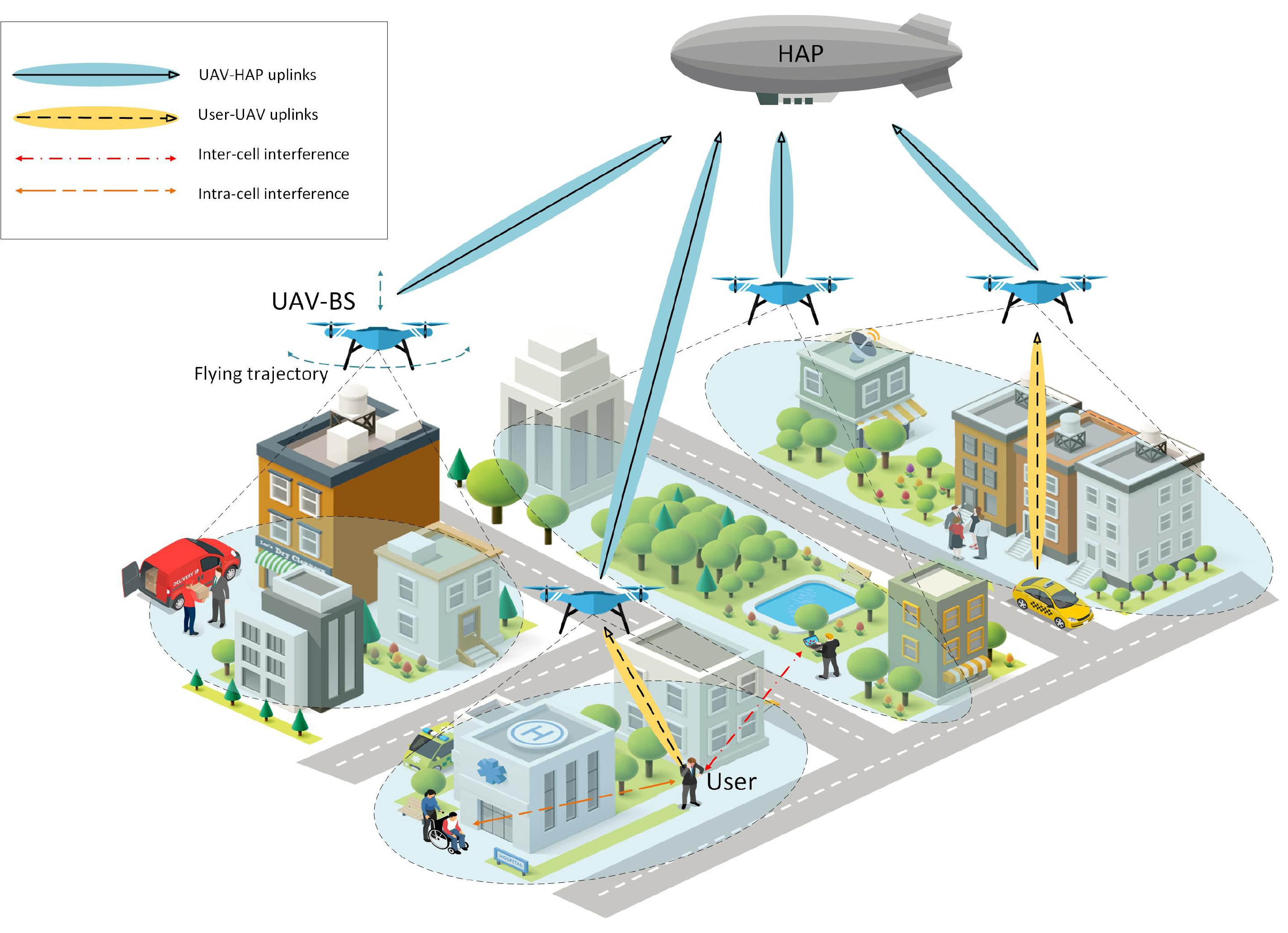}
	\caption{System model: Hierarchical NTN, user-UAV and UAV-HAP links.}
	\label{UAVpic}
\end{figure*}
\indent In the users' side, we assume that if they have any update at time slot $t$, they transmit their task to the assigned UAV-BSs via wireless link. We assume that each user has a set of tasks $\mathcal{S}$ to be sent and processed, in which $\mathcal{S}=\{1, \dots, S\}$, where $S$ is the maximum number of tasks. Let $I_{m,s}(t)$ be the indicator of whether user $m$ has task $s$ to send $I_{m,s}(t) =1$ or not $I_{m,s}(t)=0$.
%\begin{align}
%	I_{m,s}(t)=	
%	\begin{cases}
%		1, & \text{If user $m$ has task $s$ in time slot $t$},
%		\\
%		0, & \text{Otherwise}.
%	\end{cases}
%\end{align}	
\subsection {Channel Model}
We consider a multi-carrier NOMA system model for transmitting data from users to UAV-BSs. If users have data, they will transmit their data updates to the UAV-BS, independently from other users, through $N$ subcarriers (SCs) at time slot $t$. We denote the indices of subcarriers by $\mathcal{N} = \{1, 2, \dots, N\}$, where $N$ is the total number of subchannels. Let $K_{m}^{u, n}(t)$ be the indicator whether UAV-BS $u$ allocates SC $n$ to user $m$ at time slot $t$, ($K_{m}^{u, n}(t)=1$) or not ($K_{m}^{u, n}(t)=0$).
%\begin{align}
%	K_{m}^{u, n}(t)=	
%	\begin{cases}
%		1, & \text{If UAV-BS $u$ allocates the SC $n$} 
%		\\ & \text{to user $m$ at time slot $t$,}
%		\\
%		0, & \text{Otherwise}.
%	\end{cases}
%\end{align}
%As we deploy OFDM, we can assume that  there is no interference between adjacent subcarriers. 
Note that at each time slot, a user can only transmit its task through one subchannel to one UAV-BS at time slot $t$, i.e. $\sum_{n \in \mathcal{N}}K_{m}^{u, n}(t) \leq 1$. Furthermore, each subchannel is allocated to a maximum of two users at time slot $t$, i.e. $\sum_{m \in \mathcal{M}}K_{m}^{u, n}(t) \leq 2$. The UAV-BS receives the superimposed message signal of different users and applies SIC to decode each signal. The channel gain from user $m$ assigned to UAV-BS $u$ on SC $n$ is denoted by
\begin{align}\label{channel gain}
	&g_{m}^{u, n}(t)= \frac{\beta_{0}}{d_{m}^{u}(t)^{2}},
	\\\nonumber
	&d_{m}^{u}(t)=\sqrt{\left(x_{u}(t)-x_{m}^{u}(t)\right)^{2}+\left(y_{u}(t)-y_{m}^{u}(t)\right)^{2}+ h_{u}(t)^{2}},
\end{align}
where $\beta_{0}$ denotes the power gain of a channel with reference distance $d_{0}=1 $ $\text{m}$ and $d_{m}^{u}(t)$ denotes the distance between the UAV-BS $u$ and user $m$ at time slot $t$. The channel gains of all users belonging to UAV-BS $u$ on SC $n$ are sorted as $|g_{1}^{u, n}|^{2} \leq \dots \leq|g_{m}^{u, n}|^{2} \leq \dots \leq |g_{M}^{u,n}|^{2}$. We assume that all users are decoded by SIC based on these levels of channel conditions, i.e., the data signals of users in the strong channel conditions must be decoded after the data signals of users in the weak channel conditions are subtracted. We define a permutation $\zeta: \mathcal{M}\rightarrow \mathcal{M}$ to denote the SIC decoding order. For example, if the UAV-BS decodes the users' signals with the order $3 \rightarrow 1 \rightarrow 2$, the permutation $\zeta$ is set to $\zeta(1)=3$, $\zeta(2) = 1$, $\zeta(3)= 2$. According to \cite{uplinkNOMA}, the data rate transmitted from the user $m$ on SC $n$ received at the UAV-BS $u$ at time slot $t$, can be expressed as
\begin{align}\label{userdatarate}
	R_{m}^{u, n}(t)= B_{n}  \log_{2}\left(1+\frac{G_{\zeta(m)}^{u,n}(t)p_{\zeta(m)}^{u,n}(t) K_{\zeta(m)}^{u,n}(t)}{I_{\zeta(m),\text{intra}}^{u,n}(t)+ I_{\zeta(m), \text{inter}}^{u, n}(t) +1}\right),
\end{align}
where the first term in denominator represents interference between users of SC $n$ of $u$-th UAV-BS and the second term denotes to interference from other UAV-BSs' users to users in SC $n$ of UAV-BS $u$, and we have 
\begin{align}
	&I_{\zeta(m), \text{intra}}^{u, n}(t) = \sum\limits_{i=m+1}^{M}K_{\zeta(i)}^{u, n}(t)G_{\zeta(i)}^{u, n}(t)p_{\zeta(i)}^{u, n}(t),
	\\\nonumber
	&I_{\zeta(m),\text{inter}}^{u, n}(t) = \sum\limits_{\substack{u^{\prime} \in \mathcal{U} \\ u^{\prime} \neq u}}\sum_{j=m+1}^{M}K_{\zeta(j)}^{u^{\prime}, n}(t)G_{\zeta(j)}^{u^{\prime},n}(t)p_{\zeta(j)}^{u^{\prime},n}(t),
\end{align}
where $B_{n}= \frac{B}{N}$, $p_{m}^{u, n}(t)$ and $G_{m}^{u, n}(t)$ represent the bandwidth of SC $n$, assigned power for the user $m$ of UAV-BS $u$ on SC $n$ at time slot $t$, and normalized channel gain to power of additive noise $G_{m}^{u, n} (t) = |g_{m}^{u,n}(t)|^{2}/\sigma_{z}^{2}$, respectively. $B$ is the bandwidth of each UAV-BS. Also, power of each user should satisfy  
\begin{equation}
	C 3: \:\sum_{n \in \mathcal{N}}K_{m}^{u, n}(t) p_{m}^{u, n}(t) \leq p^{\text{max}}, \forall u \in \mathcal{U}, \forall m \in \mathcal{M}, 
\end{equation} 
where $p^{\text{max}}$ is the maximum power constraint for each user.  
It is assumed that the UAV-BS that receives the user's task will process only as much as it has been allocated computing capacity for, and will send the remainder to the HAP; and since this is not possible within the same time slot as the task is received, it is stored in a memory in order to be sent in the next time slots. Hence, the binary variable for sending the task of user $m$ that remained in UAV-BS $u$ to the HAP at time slot $t$ is set to $\phi_{m, s}^{u, \text{HAP}}(t)=1$ in this case, and $\phi_{m, s}^{u, \text{HAP}}(t)=0$, otherwise.
%\begin{align}
%			\phi_{m, s}^{u, \text{HAP}}(t)=	
%			\begin{cases}
%				1, & \text{If UAV-BS $u$ forwards the task $s$ of user} 
%				\\ & \text{$m$ to the HAP at time slot $t$,}
%				\\
%				0, & \text{Otherwise}.
%			\end{cases}
%\end{align}
The transmission rate of the UAV-BS $u$ to the HAP at time slot $t$ is formulated as follows:
$$
\resizebox{\linewidth}{!}{
	\begin{minipage}{1.1\linewidth}
		\begin{align}\label{rateU2U}
			&R^{u,\text{HAP}}(t)= B^{u, \text{HAP}}\log_{2}\left(1+ \frac{p^{u,\text{HAP}}(t)G^{u,\text{HAP}}L_{s}L_{l}\phi_{m,s}^{u,\text{HAP}}(t)}{k_{B}T_{\text{temp}}B^{u, \text{HAP}}}\right),
			\\ \nonumber
			&\forall u \in \mathcal{U}, m \in \mathcal{M}, t\in \mathcal{T},
		\end{align}
	\end{minipage}
}
$$
where $B^{u, \text{HAP}}$ and $G^{u, \text{HAP}}$ are bandwidth of UAV-HAP channel and antenna power gain, respectively. Also, $L_{l}$ is the total line loss, and $L_{s}=\left(\frac{c}{4\pi d^{u, \text{HAP}}(t)f^{u, \text{HAP}}}\right)^{2}$ is the free space path loss. Furthermore, $k_B$, $f^{u, \text{HAP}}$, and $T_{\text{temp}}$ denote the Boltzmann's constant, center frequency, and the system noise temperature, respectively. Wherein, $c$ is the speed of light, $d^{u, \text{HAP}}(t)$ is the distance between UAV-BS $u$ and the HAP calculated by (\ref{UAVHAPdis}) below, and $f^{u, \text{HAP}}$ is the center frequency:
$$
\resizebox{\linewidth}{!}{
	\begin{minipage}{1.2\linewidth}
		\begin{align}\label{UAVHAPdis}
			&d^{u, \text{HAP}}(t) = \sqrt{\left(x_{u}(t)-x_{\text{HAP}}\right)^{2}+\left(y_{u}(t)-y_{\text{HAP}}\right)^{2}+ \left(h_{u}(t)-h_{\text{HAP}}\right)^{2}},
			\\\nonumber
			& \forall u \in \mathcal{U}, t\in \mathcal{T}.\\\nonumber
		\end{align}
	\end{minipage}
}
$$
%\resizebox{\linewidth}{!}{
%	\begin{minipage}{1.1\linewidth}
%		\begin{align}\label{UAVHAPdis}
%			&d^{u, \text{HAP}}(t) = \sqrt{\left(x_{u}(t)-x_{\text{HAP}}\right)^{2}+\left(y_{u}(t)-y_{\text{HAP}}\right)^{2}+ \left(h_{u}(t)-h_{\text{HAP}}\right)^{2}}
%			\\\nonumber
%			& \forall u \in \mathcal{U}, t\in \mathcal{T},\\\nonumber
%		\end{align}
%	\end{minipage}
%}
 $p^{u, \text{HAP}}(t)$ is the transmit power of UAV-BS $u$. Transmit power of each UAV-BS should satisfy
\begin{align}
	&C 4: \:\phi_{m,s}^{u, \text{HAP}}(t) p^{u, \text{HAP}}(t) \leq p^{u,\text{max}},
	\\\nonumber
	&  \forall u \in \mathcal{U}, m \in \mathcal{M}, s \in \mathcal{S}, t\in \mathcal{T},
\end{align}
where $p^{u,\text{max}}$ is the maximum transmit power of each UAV-BS.
\subsection{Computing Capacity Allocation}
If UAV-BS $u$ has sufficient computing capacity, the transmitted part of task $s$ of user $m$ will be executed. Therefore, we assign a variable $\theta_{m,s}^{u}(t)\in\left[0, 1\right]$ to capture the amount of allocated computing capacity to task $s$ of user $m$ at UAV-Bs $u$ at time slot $t$. We denote $C_{m,s}^{u}(t)$ to be the computing capacity of the UAV-BS $u$ allocated to compute the task $s$ of user $m$ at time slot $t$ that can be calculated as follows:
\begin {align}
&C_{m,s}^{u}(t) = \theta_{m,s}^{u}(t) c^{u}\alpha_{m, s}(t) , \forall u \in \mathcal{U}, m \in \mathcal{M},  s \in \mathcal{S}, t\in \mathcal{T},
\end {align}
where $\alpha_{m,s}(t)$ and $c^{u}$ are the fraction of tasks sent from the task $s$ of the user $m$, which is equal to data rate of that user in time slot $t$, and the computing capacity of one bit of data. Our assumption is that the size of computation tasks is fixed and the same for all users and it is denoted by $\alpha_{m,s}$. Taking into account the maximum capacity available in each UAV-BS, $C_{u}^{\text{max}}$, the following requirement must hold
 $$
\resizebox{\linewidth}{!}{
	\begin{minipage}{\linewidth}
		\begin{align}\label{capusage}
			&C 5: \:C_{\text{usage}}^{u}(t)=\sum_{m \in \mathcal{M}}\sum_{s \in \mathcal{S}} C_{m, s}^{u}(t), \: C_{\text{usage}}^{u}(t)\leq C_{u}^{\text{max}},
			\\\nonumber
			& \forall u \in \mathcal{U}, t\in \mathcal{T}.\\\nonumber
		\end{align}
	\end{minipage}
}
$$

On the HAP side, we consider $\eta_{m,s}(t)\in \left[0, 1\right]$ to be the amount of allocated computing capacity to task $s$ of user $m$.
%\begin{align}
%	\eta_{m,s}(t)=	
%	\begin{cases}
%		1, & \text{If task $s$ of user $m$ is computed} 
%		\\ & \text{at the HAP at time slot $t$,}
%		\\
%		0, & \text{Otherwise}.
%	\end{cases}
%\end{align}	
By considering $C_{m,s}^{\text{HAP}}(t)$ to be the allocated computing capacity of the HAP to task $s$ of user $m$ at time slot $t$, we have
\begin{align}
	&C_{m,s}^{\text{HAP}}(t) = \eta_{m,s}(t)c^{\text{HAP}}\beta_{m,s}^{u}(t), \forall u \in \mathcal{U}, m \in \mathcal{M}, s \in \mathcal{S}, t\in \mathcal{T},
\end{align}
where $\beta_{m,s}^{u}(t)$ is the amount of task $s$ of user $m$ sent from UAV-BS $u$ to the HAP, which is equivalent to the amount of data rate at time slot $t$, and $c^{\text{HAP}}$ represents the computing capacity of one bit of data in the HAP. Taking into account the maximum capacity available in HAP, $C_{\text{HAP}}^{\text{max}}$, the following requirement must be satisfied
\begin{align}\label{hapcpureq}
	C 6: \:& C_{\text{usage}}^{\text{HAP}}(t)=\sum_{m \in \mathcal{M}}\sum_{s \in \mathcal{S}} C_{m,s}^{\text{HAP}}(t), \: C_{\text{usage}}^{\text{HAP}}(t)\leq C_{\text{HAP}}^{\text{max}}, \forall t\in \mathcal{T}.
\end{align}
\subsection{Remaining Tasks of the User, the UAV-BS and the HAP}
\subsubsection{User Side}
Assuming that all tasks can take more than one time slot to be transmitted, we calculate the remaining task of each user in user side at each time slot $\alpha_{m,s}^{\text{rem}}(t)$, and it can be represented as
 $$
\resizebox{\linewidth}{!}{
	\begin{minipage}{0.9\linewidth}
		\begin{align}\label{remuser}
			&\alpha_{m,s}^{\text{rem}}(t) = \alpha_{m,s}^{\text{rem}}(t-1) - \alpha_{m,s}(t),\:\alpha_{m,s}^{\text{rem}}(0) = \alpha_{m,s},
			\\\nonumber
			& \forall m \in \mathcal{M}, s \in \mathcal{S}, t\in \mathcal{T}.\\\nonumber
		\end{align}
	\end{minipage}
}
$$
\subsubsection{UAV-BS Side}
The remaining tasks in UAV-BS $u$ at time slot $t$ can be expressed as
%$$
%\resizebox{\linewidth}{!}{
%	\begin{minipage}{0.9\linewidth}
%		\begin{align}\label{remtask1}
%			O_{m, s}^{u}(t) &= (1-\theta_{m, s}^{u}(t))K_{m}^{u}(t)I_{m,s}(t)\alpha_{m, s}(t)
%			\\\nonumber
%			&+\sum_{\tau = 1}^{t-1}(1-\phi_{m,s}^{u, \text{HAP}}(t-\tau))(1-\theta_{m, s}^{u}(t-\tau)) K_{m}^{u}(t-\tau)I_{m,s}(t-\tau)\alpha_{m, s}(t-\tau)
%			\\\nonumber
%			&-\phi_{m,s}^{u, \text{HAP}}(t)\beta_{m, s}^{u}(t), \forall u \in \mathcal{U}, m \in \mathcal{M}, s \in \mathcal{S}, t\in \mathcal{T},
%		\end{align}
%	\end{minipage}
%}
%$$
\begin{align}\label{remtask1}
	O_{m, s}^{u}(t) &= (1-\theta_{m, s}^{u}(t))K_{m}^{u}(t)I_{m,s}(t)\alpha_{m, s}(t)
	\\\nonumber
	&+\sum_{\tau = 1}^{t-1}(1-\phi_{m,s}^{u, \text{HAP}}(t-\tau))(1-\theta_{m, s}^{u}(t-\tau)) 
	\\\nonumber
	& \qquad \quad K_{m}^{u}(t-\tau)I_{m,s}(t-\tau)\alpha_{m, s}(t-\tau)
	\\\nonumber
	&-\phi_{m,s}^{u, \text{HAP}}(t)\beta_{m, s}^{u}(t), \forall u \in \mathcal{U}, m \in \mathcal{M}, s \in \mathcal{S}, t\in \mathcal{T},
\end{align}
where $K_{m}^{u}(t)$ denotes to whether UAV-BS $u$ allocates a subchannel to user $m$ ($K_{m}^{u}(t)=1$) or not ($K_{m}^{u}(t)=0$), and $O_{m, s}^{u}(t)$ represents the fraction of task $s$ of user $m$ located in UAV-BS $u$. It should be mentioned that different parts of a task from one user may be sent to different UAV-BSs. In (\ref{remtask1}), the first term indicates the amount of tasks that are sent to the UAV-BS in time slot $t$, but the computing capacity is not sufficient to process them. The second term indicates the amount of tasks that were sent in the previous time slots to UAV-BS and, in addition to not being processed, they were not sent to the HAP either. Finally, the third term is the amount of tasks that are sent to the HAP at time slot $t$. Therefore, to calculate the remaining tasks of users in UAV-BS $u$, we have
\begin{align}\label{remtask2}
	O^{u}(t) = \sum_{m \in \mathcal{M}}\sum_{s \in \mathcal{S}} O_{m, s}^{u}(t), \forall u \in \mathcal{U}, t\in \mathcal{T}.
\end{align}
\subsubsection{HAP Side} 
\begin{figure}[]
	\centering
	\captionsetup{justification=centering}
	\includegraphics[width=0.5\textwidth]{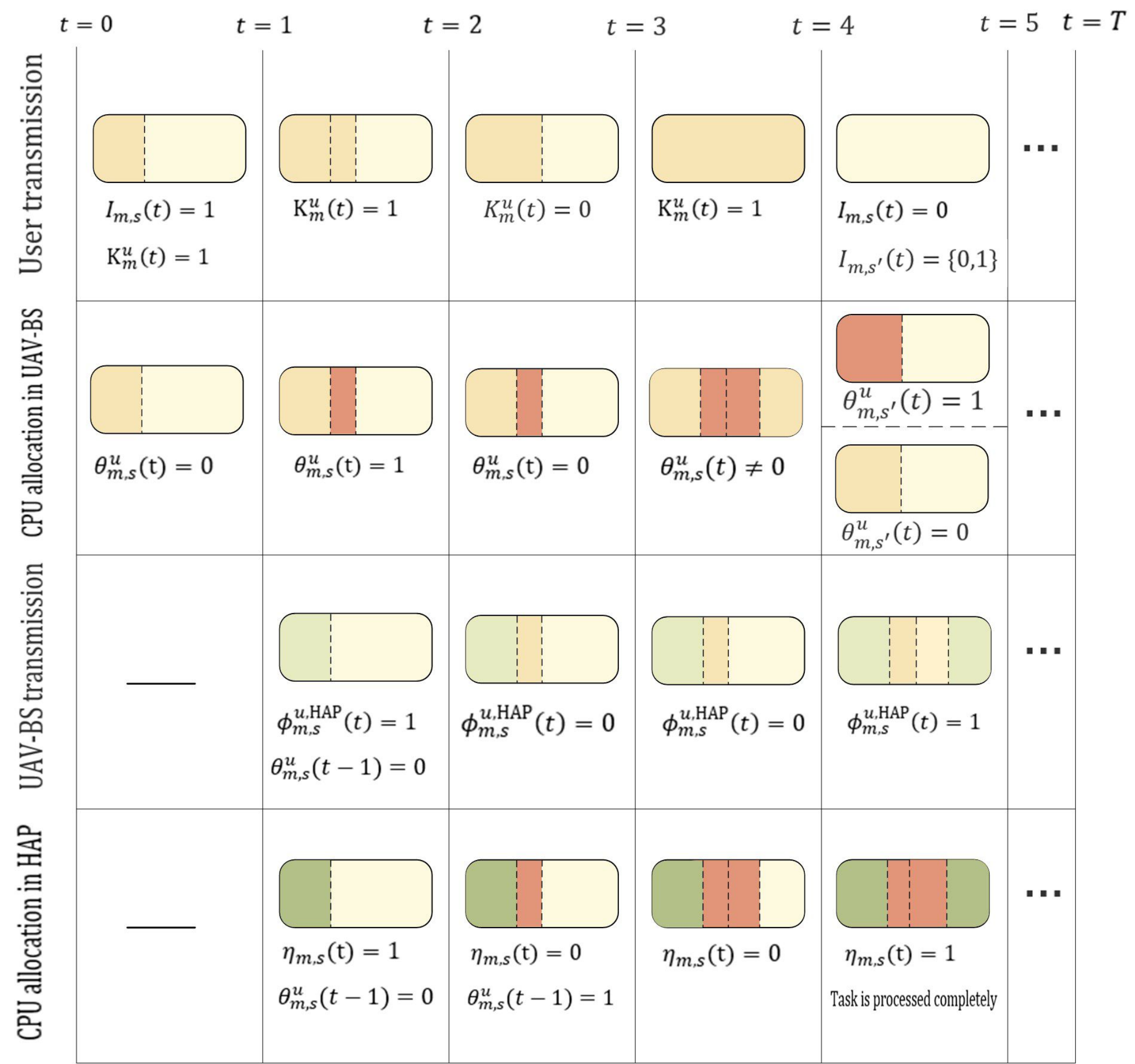}
	\caption{An example of the system operation.}
	\label{sample}
\end{figure}

The remaining tasks of users in the HAP at time slot $t$ can be calculated as 
%$$
%\resizebox{\linewidth}{!}{
%	\begin{minipage}{1.15\linewidth}
%		\begin{equation}
%			\begin{aligned}\label{remtask3}
%				O^{\text{HAP}}(t) &= \sum_{u \in \mathcal{U}} \sum_{m \in \mathcal{M}}\sum_{s \in \mathcal{S}} (1-\eta_{m,s}(t))\phi_{m,s}^{u, \text{HAP}}(t)\beta_{m, s}^{u}(t)
%				\\ 
%				&+\sum_{\tau=1}^{t-1}\sum_{u \in \mathcal{U}} \sum_{m \in \mathcal{M}}\sum_{s \in \mathcal{S}}(1-\eta_{m,s}(t-\tau))\phi_{m,s}^{u, \text{HAP}}(t-\tau)\beta_{m, s}^{u}(t-\tau), 	
%			\end{aligned}
%		\end{equation}
%	\end{minipage}
%}
%$$
%\begin{equation}
%	\begin{aligned}\label{remtask3}
%		O^{\text{HAP}}(t) &= \sum_{u \in \mathcal{U}} \sum_{m \in \mathcal{M}}\sum_{s \in \mathcal{S}} (1-\eta_{m,s}(t))\phi_{m,s}^{u, \text{HAP}}(t)\beta_{m, s}^{u}(t)
%		\\ 
%		&+\sum_{\tau=1}^{t-1}\sum_{u \in \mathcal{U}} \sum_{m \in \mathcal{M}}\sum_{s \in \mathcal{S}}(1-\eta_{m,s}(t-\tau))
%		\\
%		& \qquad \qquad \qquad \qquad \phi_{m,s}^{u, \text{HAP}}(t-\tau)\beta_{m, s}^{u}(t-\tau), 	
%	\end{aligned}
%\end{equation}
$$
\resizebox{\linewidth}{!}{
	\begin{minipage}{1.2\linewidth}
		\begin{align}
			O^{\text{HAP}}(t) = \sum_{\tau=0}^{t-1}\sum_{u \in \mathcal{U}} \sum_{m \in \mathcal{M}}\sum_{s \in \mathcal{S}}(1-\eta_{m,s}(t-\tau)) \phi_{m,s}^{u, \text{HAP}}(t-\tau)\beta_{m, s}^{u}(t-\tau), 	
		\end{align}
	\end{minipage}
}
$$
in which it indicates the tasks that are entered into the HAP at time slot $t$ and earlier times, but there is not enough capacity to process them.\\
\indent In Fig. \ref{sample}, an example of system operation for task $s$ of user $m$ is demonstrated, in which the X-axis refers to the time slot and the Y-axis refers to the user/UAV-BS transmission and CPU allocation in UAV-BSs and the HAP. A portion of the task is received in the UAV-BS during the first time slot (in dark yellow), but there is no computing capacity available. It must be sent to the HAP in the next time slot. During the second time slot, another part of the task is sent to the UAV-BS, which may be different from the previous UAV-BS. Computing capacity is sufficient for this part of the task (in red). The remaining task from the previous time slot is forwarded to the HAP (in light green) and capacity is allocated (in dark green). In the third time slot, no channel is assigned to the user. During the fourth time slot, however, the channel is assigned to the user, and computing capacity is allocated to a portion of the transmitted task, with the remainder sent to the HAP. The next task $s^{\prime}$of the user can be generated in the fifth time slot, while the remaining task from the UAV-BS is forwarded to the HAP to be processed. In the end, the whole task is processed, partially in UAV-BSs (in red) and partially in the HAP (in dark green).
\subsection {Age of Information}
Due to the importance of freshness of data at receiver side, we use the AoI matrix to measure information timeliness. Let $F_{m,s}$ be the time at which the most up-to-date task $s$ was generated in user $m$

\begin{subequations}\label{opt}
	\begin{align}
		&A_{m}(t) =
		\begin{cases}
			t-F_{m, s},  &\text{if}\quad(\ref{aooi}),\\
			A_{m}(t-1)+1,  &\text{otherwise,}
		\end{cases}\\
		&\sum_{\tau=0}^{t} \sum_{u \in \mathcal{U}}C_{m,s}^{u}(\tau)+\sum_{\tau=0}^{t}C_{m,s}^{\text{HAP}}(\tau)\geq \alpha_{m, s}, \forall m \in \mathcal{M}, s \in \mathcal{S},\label{aooi}
	\end{align}
\end{subequations}
where $A_{m}(t)$ is the AoI of user $m$ at time slot $t$, which we interpret it as follows: as long as a task has not been fully received and processed, the AoI of user increases by one for each time slot, and the value of the AoI of a user equals to the time elapsed since receiving the last task of that user.  We are able to calculate the AoI of the user based on (\ref{aooi}), which indicates that all the allocated computing capacity on both UAV-BS and HAP sides must be greater than or equal to the task size. To be more specific, there are three cases of increasing the AoI for each user. In the first case, the user has a task to send but UAV-BS does not allocate any subchannel for the task; in the second case, the user has a task and a channel for sending it, but the total allocated computing capacity is less than the user's task; and in the third case, the user sends the part or all of a task to HAP through UAV-BS but the total allocated computing capacity in both sides of UAV-BSs and HAP is less than the user's task. 
\subsection {Stochastic Channel Uncertainty}
Owing to the inherent random nature of wireless channels, it is impossible to obtain perfect CSI. As such, channel uncertainties are considered in this subsection. Assuming the channel estimation error follows the Gaussian CSI error model \cite{uncertainty1,uncertainty2, uncertainty3}, we have
\begin{equation}\label{uncertainty of channel gains}
	\mathcal{R}_{g}=\{G_{m}^{u,n}|\hat{G}_{m}^{u, n}+\Delta G_{m}^{u,n}, \Delta G_{m}^{u,n}\sim \mathcal{CN}(0,\sigma^{2}_{e})\},
\end{equation}
where $\hat{G}_{m}^{u,n}$ denotes channel estimation value that are known at the transmitters by channel estimation algorithms and channel feedback \cite{chunceres} and $\Delta G_{m}^{u,n}$ corresponds to estimation error with the variances $\sigma^{2}_{e}$. 
%Hence, under imperfect CSI (\ref{userdatarate}) becomes
%\begin{equation}\label{uncertaindatarate}
%	\resizebox{0.5\textwidth}{!}{$R_{k_{n}}(t)=\omega B \log_{2}(1+\frac{P_{k_{n},n}(t)g_{k_{n},n}(t)I_{k_{n}}(t)}{\sum\limits_{n^{\prime}=n+1}^{K_{n}}P_{k_{n^{\prime}},n}(t)g_{k_{n^{\prime}},n}(t)I_{k_{n^{\prime}}}(t)+\sum\limits_{\substack{m\neq n \\ k_{m}\in K}}P_{k_{m},n}(t)g_{k_{m},n}(t)I_{k_{m}}(t) +N_{noise}B\omega})$}	\end{equation}	
To ensure minimum SINR requirement of each user, we put a constraint on data rate:
\begin{align}\label{SINR requirement}
	&R_{m}^{u,n}(t)\geq R_{\text{min}}, \forall u \in \mathcal{U}, m \in \mathcal{M}, \\\nonumber
	&n \in \mathcal{N}, t=\{t| K_{m}^{u}(t)=1, t \in \mathcal{T}\},
\end{align}
where $R_{\text{min}}$ is minimum rate requirement of all users. We can limit the actual outage rate probability of each user within a threshold value and achieve a trade-off between optimality and robustness	
\begin{align}\label{outage rate}
	C 7: \:&\textrm{Pr}[R_{m}^{u,n}(t)\leq R_{\text{min}}|\Delta G_{m}^{u,n}\in \mathcal{R}_{g}]\leq \epsilon_{m},
	\\\nonumber
	&\forall u \in \mathcal{U}, m \in \mathcal{M}, n \in \mathcal{N}, t=\{t| K_{m}^{u}(t)=1, t \in \mathcal{T}\}.
\end{align}	
\subsection{Problem Formulation}\label{III}
The objective of this paper is to minimize the AoI of all users subject to UAV-BS trajectory and power and computing capacity allocation in NTN under CSI uncertainty constraint. Our problem for $t \in \mathcal{T}$ can be formulated as follows:
\begin{subequations}\label{opt}
	\begin{align}
		\min\limits_{\substack{\textbf{v}_{\mathcal{U}},\textbf{K}, 	\textbf{P}_{\mathcal{M}}, \\ \boldsymbol{\theta}, \textbf{P}_{\mathcal{U}}, \boldsymbol{\eta}}} &  \Big\{A_{\mathcal{M}}^{\mathcal{T}} = \sum_{t=0}^{T}\sum_{m=0}^{M} A_{m}(t)\Big\}\\
		\text { s.t. } \quad &C 1-C 7, \\
		\quad &\sum_{n \in \mathcal{N}}K_{m}^{u, n}(t) \leq 1,  \forall u \in \mathcal{U}, \forall m \in \mathcal{M},  t \in \mathcal{T}, \\  
		\quad &\sum_{m \in \mathcal{M}}K_{m}^{u, n}(t) \leq 2,  \forall u \in \mathcal{U}, \forall n \in \mathcal{N}, t \in \mathcal{T}, \\
		\quad &h_{\textrm{min}}\leq h_{u}(t) \leq h_{\textrm{max}},  	\forall u \in \mathcal{U},  t \in \mathcal{T}, \\
		\quad &v_{u}(t) \leq v_{\text{max}},  \forall u \in \mathcal{U},  t \in \mathcal{T},
	\end{align}
\end{subequations}
where $\textbf{v}_{\mathcal{U}}=\{v_{u}(t), u \in \mathcal{U}, \}$, $\textbf{K}=\{K_{m}^{u,n}(t), u \in \mathcal{U}, m \in \mathcal{M}, n \in \mathcal{N}\}$ and $\textbf{P}_{\mathcal{M}}=\{p_{m}^{u,n}(t), u \in \mathcal{U}, m \in \mathcal{M}, n \in \mathcal{N}\}$ denote the velocity of all UAV-BSs, channel allocation between users and UAV-BS and transmission power allocation of all users, respectively. $\boldsymbol{\theta} = \{\theta_{m, s}^{u}(t), u \in \mathcal{U}, m \in \mathcal{M}, s \in \mathcal{S}\}$ denotes the CPU allocation to tasks of users in all UAV-BSs. Moreover, $\textbf{P}_{\mathcal{U}}=\{p^{u, \text{HAP}}(t), u \in \mathcal{U}\}$ and $\boldsymbol{\eta} = \{\eta_{m,s}(t), m \in \mathcal{M}, s \in \mathcal{S}\}$, denote the transmission power of all UAV-BSs and computing capacity allocation in the HAP, respectively.
\section{MADDPG and FRL Approaches to Solve the Problem} \label{IV}
The optimization problem (\ref{opt}) is a mixed integer non-linear programming (MINLP) problem, which is non-convex and difficult to solve globally. In this section, we adopt online learning based algorithms to handle the resource allocation and trajectory planning. To do so, we have to reformulate it as Markov Decision Process (MDP) and then apply two DRL-based algorithms to solve it.
\subsection{Markov Decision Process}
In this problem we deal with two kinds of agents, a HAP and multiple UAV-BSs, as they choose their own actions. However, both types of agents follow the same goals, which are minimizing average AoI. Agents gradually learn the model-free policy that maps observations into optimal actions. In the DRL framework, MDP is a tuple $\{s_{t}, a_{t}, r_{t}, s_{t+1} \}$ containing states, actions, rewards, and new states:\\
\indent \textbf{States} ${s}_{\text{s}, t}= \{s_{\text{s},t}^{u}, s_{\text{s},t}^{\text{HAP}}\}, {s}_{\text{s}, t} \in \mathcal{S}_{\text{s}}, u \in \mathcal{U} $: At each time slot $t$, the set of states of UAV-BS $u$ is expressed as
\begin{align} \label{uavstate}
	s_{\text{s},t}^{u} = \{\mathbf{q}_{u}(t), \mathbf{v}_{m}(t), I_{m,s}(t), C_{\text{usage}}^{u}(t), \alpha_{m,s}^{\text{rem}}(t), O^{u}(t), A_{m}(t) \}.
\end{align}
\indent The states of UAV-BS $u$ include location of UAV-BS and users, binary variable of existing task in network, amount of CPU usage, remaining tasks on both user and UAV-BS sides, and also the AoI values of all users. At each time slot $t$, the set of states of HAP is defined as
\begin{align}\label{hapstate}
	s_{\text{s},t}^{\text{HAP}} = \{\mathbf{q}_{\mathcal{U}}(t), I_{m,s}(t), C_{\text{usage}}^{\text{HAP}}(t), O^{\text{HAP}}(t), O^{\mathcal{U}}(t-1), {A}_{m}(t) \}.
\end{align}
\indent The states of HAP include location of all UAV-BSs, binary variable of existing task in network, amount of CPU usage in HAP, remaining tasks in HAP side, and remaining tasks of previous time slots of all UAV-BSs, and also the AoI values of all users.\\
\indent \textbf{Actions} $a_{t} = \{a_{t}^{U}, a_{t}^{\text{HAP}}\}, a_{t} \in \mathcal{A}, u \in \mathcal{U}$: At each time slot $t$, the actions that are taken by each of UAV-BS and HAP are
\begin{align}\label{uavhapaction}
	&a_{t}^{u}=\{v_{u}(t), K_{m}^{u,n}(t), p_{m}^{u, n}(t), \theta_{m, s}^{u}(t)\},\\
	& a_{t}^{\text{HAP}} = \{p^{u ,\text{HAP}}(t), \eta_{m,s}(t)\}.
\end{align}
\indent The actions of each UAV-BS in each time slot comprise taking velocity in each direction which results in trajectory decision, subchannel allocation and power allocation to users, and CPU allocation to task of all users. The actions of the HAP comprise power allocation to UAV-BSs and CPU allocation to task of all users. Since we clip the action values in $\left[-1, 1\right]$, for integer actions we denote $1$ for action value in $\left[0, 1\right]$ and $0$ for action value in $\left[-1, 0\right)$. \\
\indent \textbf{Reward} $r_{t}$: To minimize the average AoI of all users, we define the reward as
\begin{align}
	r_{t} = -\big\{\frac{1}{M}\sum_{m} A_{m}(t)\big\} , \forall m \in \mathcal{M}.
\end{align}
\indent \textbf{Discount Factor} $\gamma$: The long-term accumulated reward of a policy is modeled as 
\begin{align}
	R_{t} = \sum_{i=0}^{\infty}\gamma^{i}r_{t+i},
\end{align}
where the discount factor $\gamma \in \left[0, 1\right]$ indicates that the agent is more concerned about the long-term reward if the discount factor approaches $1$. In addition, the transition function $T(s_{\text{s}, t}, a_{t}, s_{\text{s}, t+1})$ identifies the probability of the next state $s_{\text{s}, t+1} \in \mathcal{S}_{\text{s}}$ given the current state $s_{\text{s}, t} \in \mathcal{S}_{\text{s}}$ and the current action $a_{t} \in \mathcal{A}$ that applies on $s_{\text{s}, t}$.
\subsection{MADDPG Approach}
This approach utilizes actor-critic networks, in which each agent is equipped with four networks: actor, target actor, critic, and target critic network. The observations and actions of all RL agents are considered for learning the actor and critic networks \cite{lillicrap2015continuous}. During action execution, each agent's actor network only considers local observations, but observations of all agents along with each agent's actions are fed into critic networks of that agent to be evaluated. Based on a consistent gradient signal, each agent learns an optimal policy through centralized training and decentralized execution. To improve the online learning efficiency, the replay buffer technique is used. However, the batch size is chosen to be as small as possible. Considering a policy $\pi$, each observation maps to an action. The value function, based on which the quality of policy is evaluated, is defined as 
\begin{align}
	Q_{\pi} = \mathbb{E} \Big[ \sum_{t=0}^{T} \gamma^{t} r(s_{t}, a_{t})\Big].
\end{align}

Through optimization of the objective function, $\pi$ is parameterized to $\pi_{\theta}$, and its expected value is maximized  $J(\theta) =  \mathbb{E}[Q_{\pi_{\theta}}(s, \pi_{\theta}(s))]$. The gradient of this objective function can be calculated by applying the deterministic policy gradient and the chain rule is applied to the expected reward from the  distribution $J$ to update the policy function $\pi$.
$$
\resizebox{\linewidth}{!}{
	\begin{minipage}{1.1\linewidth}
		\begin{align}
			\nabla_{\theta^\pi} J & \approx \mathbb{E}_{s_t \sim \rho^\beta}\left[\left.\nabla_{\theta^\pi} Q\left(s, a \mid \theta^Q\right)\right|_{s=s_t, a=\pi\left(S_t \mid \theta^\pi\right)}\right], \\\nonumber
			&=\mathbb{E}_{s_t \sim \rho^\beta}\left[\left.\left.\nabla_a Q(s, a \mid \theta^Q)\right|_{s=s_t, a=\pi\left(s_t\right)} \nabla_{\theta_\pi} \pi\left(s \mid \theta^\pi\right)\right|_{s=s_t}\right] ,
		\end{align}
	\end{minipage}
}
$$
where $\rho^\beta$ is the distribution of state-visitation associated with a policy. By copying directly (hard update) or exponentially decaying average (soft update), defined by hyper parameter $\tau\ll1$, the target network is synchronized with the primary network every $T_{\text{up}}$ steps. In Algorithm \ref{alg:VFRL_MADDPG}, the process of learning in this approach is presented, with considering FRL-mode to be False.
\subsection{VFRL Approach}
In the proposed model, we are dealing with two types of agents and although they are in the same environment, they have different interactions with environment. This leads us to apply the vertical category of FRL algorithm (VFRL) to our problem \cite{qi2021federated}. Due to the decentralized nature of the model, a peer-to-peer (P2P) method has been suggested \cite{maenpaa2021towards}, in which there is no central aggregator to form a global model for all agents and they exchange their parameters among themselves. As a result, we assume that at every $T_{\text{FL}}$ steps, each agent trains the model on its local data, resulting in a model with parameter $\theta$ for the actor network. Then, each agent aggregates and averages updates from all other agents. Next, the local model parameters are updated by
\begin{align}
	\theta^{i}_{t} = \omega^{i}\theta^{i}_{t} + \sum_{i^{\prime}}\omega^{i^{\prime}} \theta^{i^{\prime}}_{t},
\end{align}
where $i$ refers to index of agents and $\omega^{i}$ is the weight of parameters of agent $i$. According to \cite{zhu2021federated}, we consider the weights as $\omega^{i}= \frac{1}{N_{\text{agent}}}$ and $\omega^{i^{\prime}}=\frac{1-\omega^{i}}{N_{\text{agent}}-1}$, where $N_{\text{agent}}$ is the number of agents. Inspired by \cite{dong2020deep, mcmahan2017communication}, we develop  Algorithm \ref{alg:VFRL_MADDPG}, which is in VFRL mode, when FRL-mode is True and MADDPG, otherwise.

	\begin{algorithm}
		\DontPrintSemicolon
		\tiny
		\scriptsize
		\KwInput{Initialize parameters $\phi$ and $\theta$ of critic and actor networks $Q$ and $\mu$, target networks $Q^{\prime}$ and $\mu^{\prime}$; set M = 0 as memory counter\\\\}
		Receive initial states\\
		\For{\textup{episode} $t=1$ to $T$}{
			%Update UAV-BSs locations and respective channel gains\\
			\For{\textup{each agent} $i \in \{1, \dots, U, H\}$}{
				Observe $s_{\text{s},t, i}$ and select action $a_{t, i}=\mu_{\theta_{i}}(s_{\text{s},t, i})$\\
				Execute actions and observe reward $r_{t, i}$ and new states $s_{\text{s},t+1,i}$\\
				Store $(\mathbf{s}_{\text{s},t}, a_{t}, r_{t}, \mathbf{s}_{\text{s},t+1})$ in reply buffer $\mathcal{D}$ and $\mathbf{s}_{\text{s},t}\leftarrow \mathbf{s}_{\text{s},t+1}$\\
				M $\leftarrow$ M$+1$\\
			}
			\If{\textup{M}$>=$\textup{B}}
			{Sample mini-batch of size D, $ (\mathbf{s}^{j}_{\text{s},t}, \mathbf{a}^j_{t}, \mathbf{r}^j_{t}, \mathbf{s}^{j}_{\text{s},t+1}) $ from replay buffer $ \mathcal{D} $\\
				\For{\textup{each agent} i}{
					Set\\ $ y^j=r^{j}+\gamma Q^{\prime}\left(\mathbf{s}^{j}_{\text{s}, t+1}, \mathbf{a}^{j}_{t+1}|\theta^{\prime}\right)\big{|}_{\mathbf{a}^{j}_{t+1}=\mu^{\prime}(\mathbf{s}^{j}_{\text{s}, t+1})} $\\
					Calculate:\\ $ \mathcal{L}=\frac{1}{D}\sum_{j}\left(y^j-Q\left(\mathbf{s}^j_{\text{s}}, \mathbf{a}^j|{\phi}\right)\right)^{2} $\\
				}
			}
			
			\If{t \textup{mod} $T_{\text{FL}}==1$ and \textup{FRL-mode}}{
				
				\For{\textup{each agent} i \textup{\textbf{in parallel}}}{
					$\theta^{i}_{t+1}$$\leftarrow$ \FMain{$i$, $\theta$, $\phi$, $\alpha$, $\beta$}\\
					$\theta^{i}_{t+1} = \omega^{i}\theta^{i}_{t+1} + \sum_{i^{\prime}}\omega^{i^{\prime}} \theta^{i^{\prime}}_{t+1}$	
				}
			}
			\Else{\For{\textup{each agent} i \textup{\textbf{in parallel}}}{
					$\theta^{i}_{t+1}$$\leftarrow$ \FMain{$i$, $\theta$, $\phi$, $\alpha$, $\beta$}\\}}

			\If{t \textup{mod} $T_{up}==1$:}{
				Update target networks parameters for each agent i: \\
				$\theta_{i}^{\prime} \leftarrow \tau\theta_{i} + (1-\tau)\theta_{i}^{\prime} $\\
				$ \phi_{i}^{\prime} \leftarrow \tau\phi_{i} + (1-\tau)\phi_{i}^{\prime} $\\
			}
		}	
		\Fn{\FMain{$i$, $\theta$, $\phi$, $\alpha$, $\beta$}}{
			Update actor using the sampled policy gradient:
			
			\resizebox{\linewidth}{!}{
				\begin{minipage}{1.2\linewidth}
					$
					\begin{aligned}
						\nabla_{\theta} J\approx
						&\frac{1}{D}\sum_{j}\Bigg\{\nabla_{\theta}\mu(s^{j}_{\text{s}})\nabla_{a} Q\left(s_{\text{s}}, a |\phi\right)|_{s_{\text{s}}=s^{j}_{\text{s}},a=\mu(s^{j}_{\text{s}})} \Bigg\}\\
					\end{aligned}
					$
				\end{minipage}
			}
			Update critic by minimizing the loss:$\mathcal{L}$\\
			
			$\theta^{i}_{t+1} = \theta^{i}_{t}-\alpha \nabla \mathcal{L}(\theta^{i}_{t})$\;
			$\phi^{i}_{t+1} = \phi^{i}_{t}-\beta \nabla \mathcal{L}(\phi^{i}_{t})$\;
			\KwRet $\theta^{i}_{t+1}$, $\phi^{i}_{t+1}$\;
		}			
		\caption{\linespread{0.5}{Online P2P-VFRL and MADDPG. The agents are indexed by $i$; B and D are batch size and local mini-batch size, respectively, and $\alpha$, $\beta$ are the learning rates of actor and critic networks.}}
		\label{alg:VFRL_MADDPG}
	\end{algorithm}
\subsection{Computational Complexity}
We analyze the computational complexity of both MADDPG and P2P-VFRL approaches. The complexity of DNN-based algorithms depends on the architecture, configuration, number of inputs and outputs, and hidden layers. Moreover, we have to take into account 
the action and state space size, number of neural networks, number of trainable variables, and the communication overhead between agents and the central servers. 
To this end, we assume $|S_{\text{s}}|$ and $|A|$ to be action and state space size, respectively; which according to (\ref{uavstate}), (\ref{hapstate}) and (\ref{uavhapaction}), are calculated as follows:
%\begin{align}
%	|S_{\text{s}}| &= \underbrace{2\times M + 4\times(M\times S)+3\times U+M}_{\text{UAV-BS states}=|S^{u}_{\text{s}}|} \\\nonumber
%	&+ \underbrace{3\times U+3\times (M\times S)+M+(U\times M\times S)}_{\text{HAP states}=|S^{\text{HAP}}_{\text{s}}|},
%\end{align}
\begin{align}
	|S_{\text{s}}| &= \underbrace{2\times M + 4\times(M\times S)+3\times U+M}_{\text{UAV-BS states}=|S^{u}_{\text{s}}|} \\\nonumber
	&+ \underbrace{3\times U+3\times (M\times S)+M+(U\times M\times S)}_{\text{HAP states}=|S^{\text{HAP}}_{\text{s}}|},\\
	|A| &=\underbrace{3 + 2\times (M\times N) + (M\times S)}_{\text{UAV-BS actions}=|A^{u}|}\\\nonumber
	& + \underbrace{(M \times U)+(M\times S)}_{\text{HAP actions}=|A^{\text{HAP}}|}.
\end{align}

Thus, the number of MADDPG parameters to train is
\begin{align}
	\mathcal{O}\big(U^{2}(|S^{u}_{\text{s}}|+|A^{u}|)+H^{2}(|S^{\text{HAP}}_{\text{s}}|+|A^{\text{HAP}}|)\big),
\end{align}
where $U$ and $H$ represent the number of UAV-BSs and HAP agents, respectively. MADDPG employs $2\times \big(N_{\text{agents}}(\b{1}_{Q}+\b{1}_{A})\big)$ neural networks as was with P2P-VFRL, where the multiplication by 2 is due to target networks, and $\b{1}_{Q}$ and $\b{1}_{A}$ represent the critic and actor networks. Moreover, $N_{\text{agents}} = U + H$ refers to number of agents. To calculate the computational complexity, first we assume that $\Gamma^{a}_{i}$, and $\Gamma^{c}_{i}$ are the number of neurons in the $i$-th layer of the actor and critic networks, respectively. We denote $L_{a}$ and $L_{c}$ to be the total number of layers in the respective actor, and critic networks. As actor and critic networks are fully interconnected, their computational complexity can be expressed as $\mathcal{O}\big(\sum_{i=2}^{i=L_{a}-1}(\Gamma^{a}_{i-1}\Gamma^{a}_{i}+\Gamma^{a}_{i}\Gamma^{a}_{i+1})\big)$, and $\mathcal{O}\big(\sum_{i=2}^{i=L_{c}-1}(\Gamma^{c}_{i-1}\Gamma^{c}_{i}+\Gamma^{c}_{i}\Gamma^{c}_{i+1})\big)$, respectively. As a result, the computational complexity of  both approaches can be obtained as
\begin{align}
	\mathcal{O}\big(N_{\text{agents}}\cdot D\cdot E_{\text{episode}}\cdot(\mathfrak{C}^{a} +\mathfrak{C}^{c})\big),
\end{align}
where $D$ and $E_{\text{episode}}$ indicate the mini-batch sampling size, and number of episodes, respectively. In addition we have $\mathfrak{C}^{a} = \sum_{i=2}^{i=L_{a}-1}(\Gamma^{a}_{i-1}\Gamma^{a}_{i}+\Gamma^{a}_{i}\Gamma^{a}_{i+1})$ and $\mathfrak{C}^{c} = \sum_{i=2}^{i=L_{c}-1}(\Gamma^{c}_{i-1}\Gamma^{c}_{i}+\Gamma^{c}_{i}\Gamma^{c}_{i+1})$. Most multi-agent reinforcement learning-based frameworks rely on agents' communication. For the learning process to be stabilized and agents to cooperate, data exchange is essential which imposes some overhead on the system. 
To obtain the communication overhead in the proposed framework, we examined how often the agents need to interact with the central server and other agents during the learning process. Accordingly, P2P-VFRL has an overhead of $N_{\text{agents}}$, while MADDPG has an overhead of $N_{\text{agents}}(N_{\text{agents}}-1)$.
\section{Numerical Results} \label{V}
\subsection{Simulation setup}
We consider one HAP with the height of 20 km, and for the initial step, $2$ UAV-BSs with an altitude of $400$ $\text{m}$ are uniformly distributed in the area with size of $200$ $\text{km}$$\times$$200$ $\text{km}$, and terrestrial mobile users are randomly distributed in this area. The task size of all users $\alpha_{m, s}$ is fixed to $10$ $\text{Mbit}$. At the beginning of the simulation, we assume that all users are assigned tasks to send. After completing each task, the next task of the user will be generated randomly with the probability of $50\%$. Furthermore, we assume that the total bandwidth of each UAV is $B = 10$ $\text{MHz}$, which is divided into $N = 8$ subchannels. The AWGN power spectrum density is $-174$ $\text{dBm/Hz}$. Each subchannel can be assigned to at most $4$ users and each user can access $2$ subchannels. The parameters associated with computation and communication are summarized in Table \ref{tabpar}. Furthermore, since users have to send high volume tasks, the maximum number of users is assumed to be $40$. %The Monte Carlo method is employed due to the randomness of task generation on users' side and channels between users and UAV-BSs based on users' movements, to get smoother results. 
Also, to deploy DNN, we use Pytorch library in Python. 
\begin{table}
	\centering	
	\caption{Simulation Parameters}
	\label{tabpar}  
	\begin{tabular}{ll}
		\hline
		\textbf{Parameter}  & \textbf{Value}            \\ \hline
		$c^{u}$             & $200\text{ cycles/bit}$          \\ 
		$c^{\text{HAP}}$    & $500 \text{ cycles/bit}$ \\ 
		$p^{u, \text{max}}$       & $0.5$ $\text{W}$          \\         
		$p^{\text{max}}$          & $0.2$ $\text{W}$            \\ 
		$k_{B}$             & $1.38\times10^{-23}$ $\text{J/K}$  \\ 
		$T_{\text{temp}}$      & $1000$ $\text{K}$              \\ 
		$G^{u, \text{HAP}}$ & $15$ $\text{dB}$                    \\ 
		$B^{u, \text{HAP}}$ & $20$ $\text{MHz}$                  \\ 
		$f^{u, \text{HAP}}$ & $2.4$ $\text{GHz}$                 \\ 
		$v_{\text{max}}$ & $50$ $\text{m/s}$                 \\ 
		$C_{u}^{\text{max}}$ & $10^{9}$ $\text{cycles}$                  \\ 
		$C_{\text{HAP}}^{\text{max}}$ & $5\times 10^{9}$ $\text{cycles}$                 \\ 
		Fast fading 	& Rayleigh fading\\\hline
		\textbf{Neural network parameter}  & \textbf{Value}            \\ \hline
		Replay buffer size & $50000$                  \\ 
		Mini batch size & $8$                  \\
		Number/size of actor networks hidden layers & $2 / 1024, 512$ \\
		Number/size of critic networks hidden layers  & $2 / 512, 256$               \\ 
		Critic/Actor networks learning rate & $0.0001/0.00001$\\
		Discount factor  &$0.99$\\
		Target networks soft update parameter, $\tau$ & $0.0005$\\
		\hline
	\end{tabular} 
\end{table}
\subsection{Simulation Results Discussions}
\begin{figure}[b]
	\centering
	\captionsetup{justification=centering}
	\includegraphics[width=0.5\textwidth]{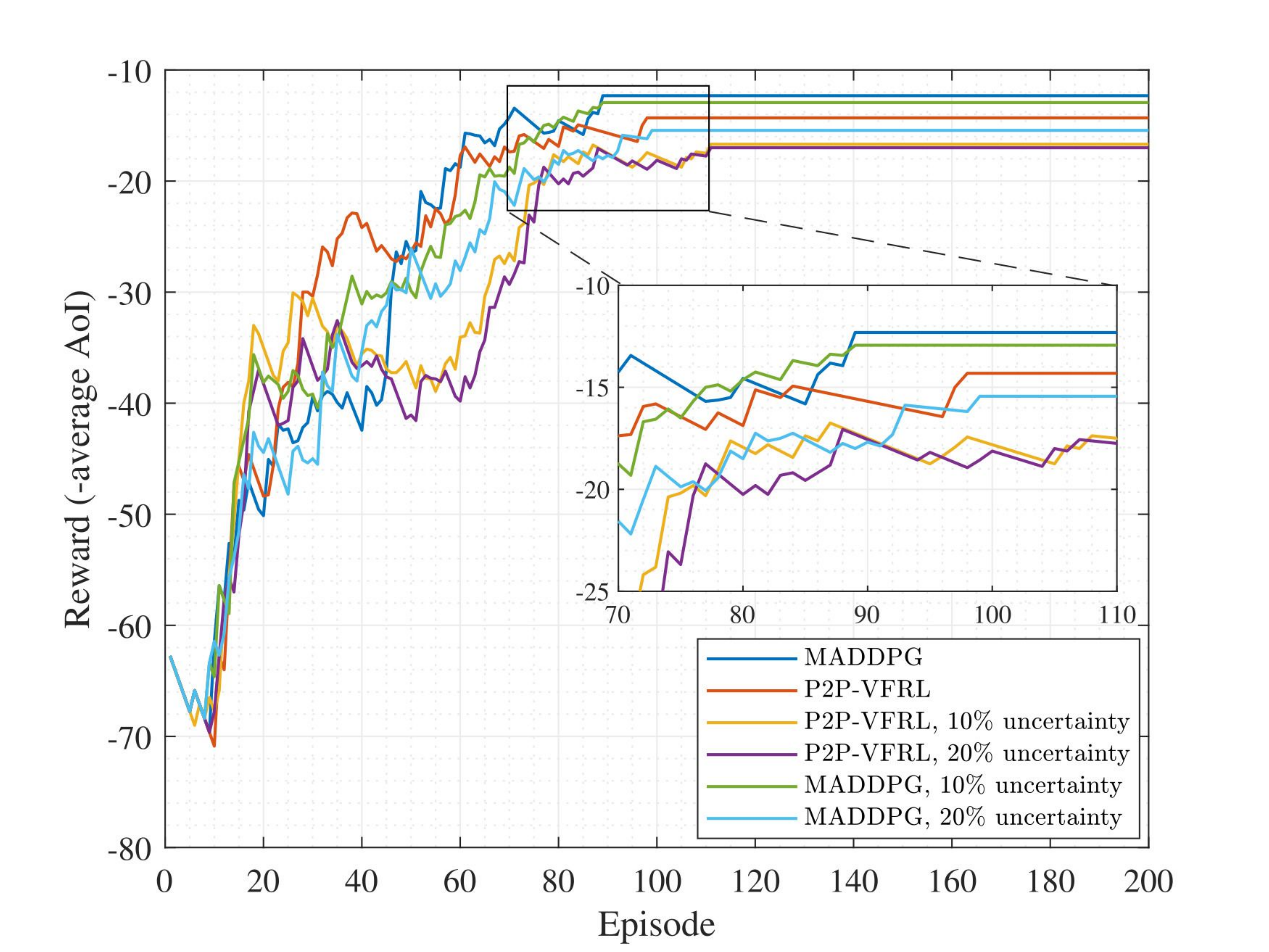}
	\caption{Numerical assessment of reward of MADDPG and P2P-VFRL approaches over learning episodes.}
	\label{reward}
\end{figure}

A comparison of the rewards of the two approaches with different levels of uncertainty (including perfect CSI) is provided in Fig. \ref{reward}, in which users send five tasks to be processed and agents allocate resources until all tasks are processed. It is observed that the MADDPG approach has better performance, as it is shown in Table \ref{tab:performancereward}. The reason is that it relies on central training and decentralized execution. After central training in MADDPG, agents may finally have different models for actor networks. On the other hand, in P2P-VFRL we deal with aggregating other agents' network parameters and updating their parameters at specific time slots and decentralized training is applied in the training phase, with decentralized execution. Given the better results achieved by MADDPG, we evaluate the effects of the main parameters with this approach including maximum available computation capacity both in UAV-BS and HAP, maximum transmission power of each user and UAV-BS, number of subchannels with considering the number of users, and uncertainty, and also the effect of task size. The source code for the
simulation of both approaches are available in \cite{8yvb-e654-22}.
\begin{table}[h]
	\centering
	\caption{Average AoI improvement between MADDPG and baselines.}
	\label{tab:performancereward}
	\begin{tabular}{|c|c|}
		\hline
		\textbf{Baselines} & \textbf{average AoI gap} \\ \hline
		P2P-VFRL           & $-16\%$                      \\ \hline
		MADDPG, $10\%$ uncertainty     & $-5\%$                     \\ \hline
		MADDPG, $20\%$ uncertainty             & $-25\%$                      \\ \hline
		P2P-VFRL, $10\%$ uncertainty           & $-35\%$                      \\ \hline
		P2P-VFRL, $20\%$ uncertainty           & $-38\%$                      \\ \hline
	\end{tabular}
\end{table}
\subsubsection{Maximum Available CPU} 
\begin{figure}
	\centering
	\captionsetup{justification=centering}
	\begin{subfigure}{0.5\textwidth}
		\captionsetup{justification=centering}
		\includegraphics[width=\textwidth]{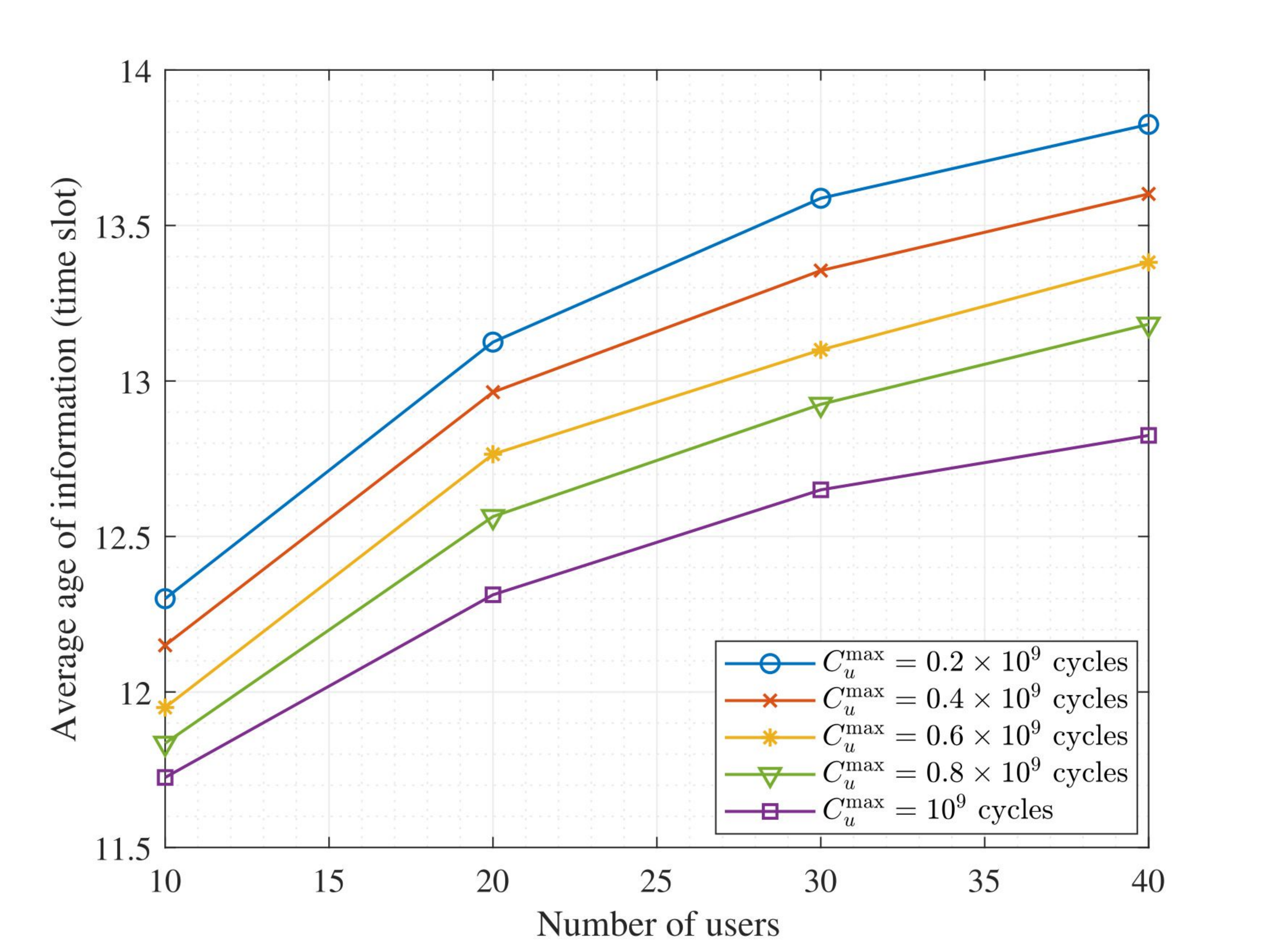}
		\caption{Impact of UAV-BSs' computation capability ($C^{\text{max}}_{\text{HAP}}= 5\times 10^{9} \text{cycles}$).}
		\label{fig:uav}
	\end{subfigure}
	\begin{subfigure}{0.5\textwidth}
		\captionsetup{justification=centering}
		\includegraphics[width=\textwidth]{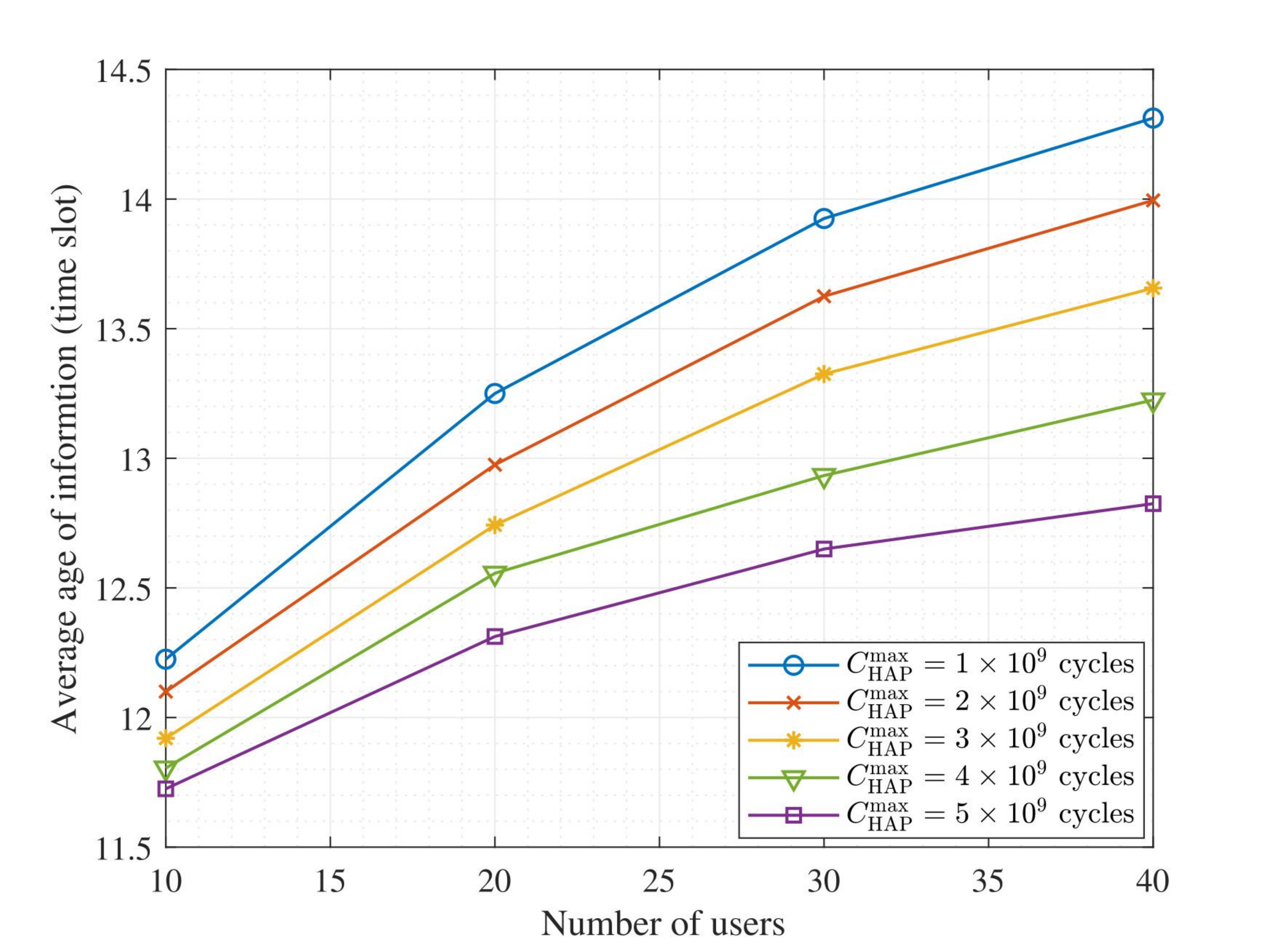}
		\caption{Impact of HAP’s computation capability ($C^{\text{max}}_{u}= 10^{9} \text{cycles}$).}
		\label{fig:HAP}
	\end{subfigure}
	\caption{Impact of computation capability on network performance.}\label{fig:CPU}
\end{figure}

To evaluate the impact of existing computation capacity in UAV-BS alongside with number of users, we consider $C^{\text{max}}_{\text{HAP}}= 5\times 10^{9}$ $\text{cycles}$ to be fixed, increase the available CPU in UAV-BSs, and plot the average AoI for different number of users in Fig. \ref{fig:uav}. As we expected, with higher amount of CPU power in UAV-BS, the tasks will be processed faster, leading to lower average AoI. Also, for analyzing the impact of available computing capacity in HAP, we consider $C^{\text{max}}_{u}= 10^{9} \text{ cycles}$ to be fixed, increase the available CPU power in the HAP, and plot the average AoI for different number of users in Fig. \ref{fig:HAP}, and a similar effect can be seen for the HAP computation capability on the average AoI.
It is observed that the variation of HAP's computation capacity has a greater impact on the optimization results compared to the variation of UAV-BS's computation capacity. This is due to the fact that we assume the UAV-BS does not decide whether to allocate CPU for the remaining tasks in UAV-BS or to send them to the HAP in the next time slot. As a result, assigning more CPU resources to the HAP will result in lower average AoI. 
\subsubsection{Maximum Transmission Power}
\begin{figure}
	\begin{subfigure}[b]{0.5\textwidth}
		\includegraphics[width=\textwidth]{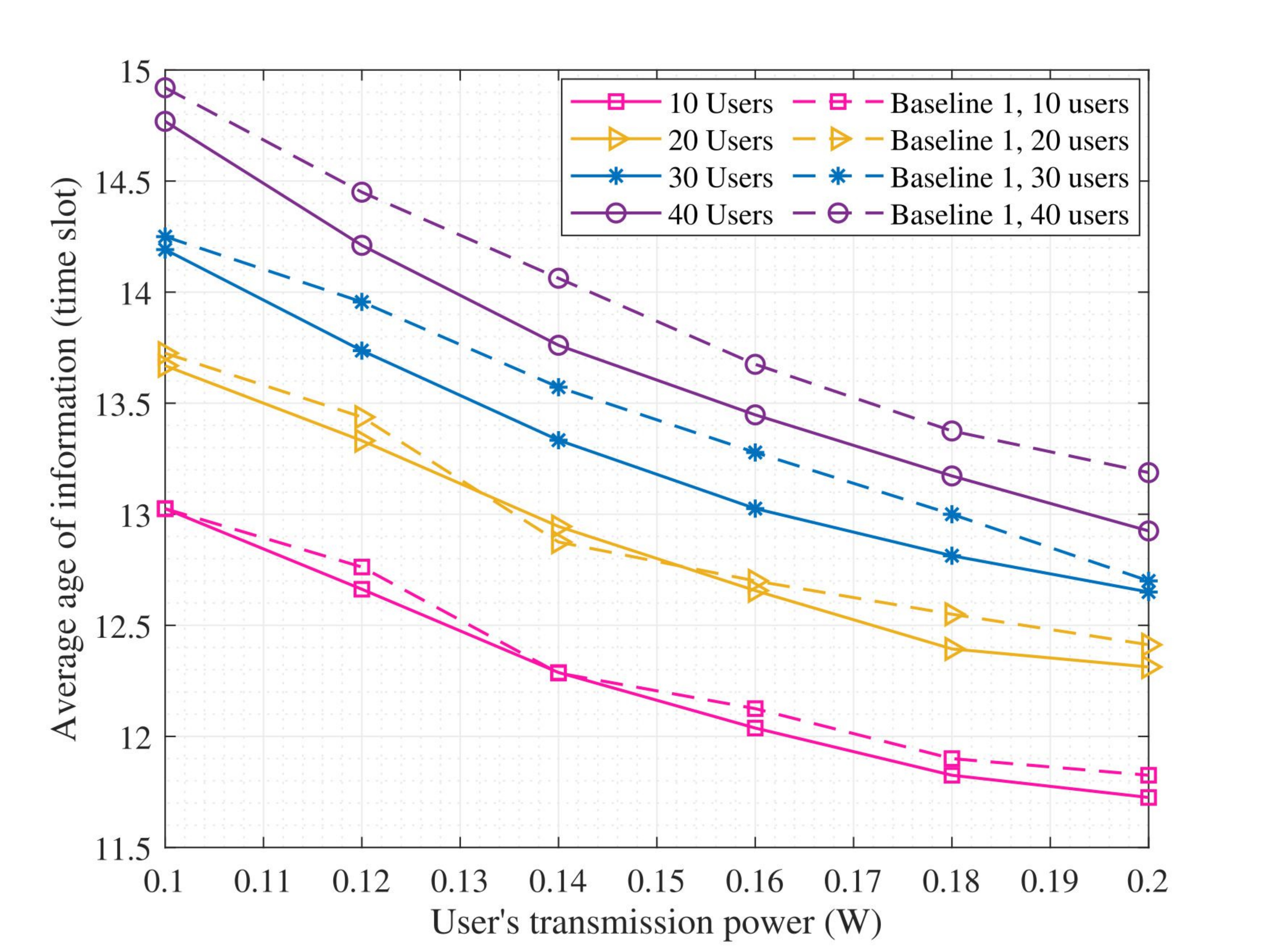}
		\caption{Impact of users' transmission power.}
		\label{fig:userp}
	\end{subfigure}
	\begin{subfigure}[b]{0.5\textwidth}
		\includegraphics[width=\textwidth]{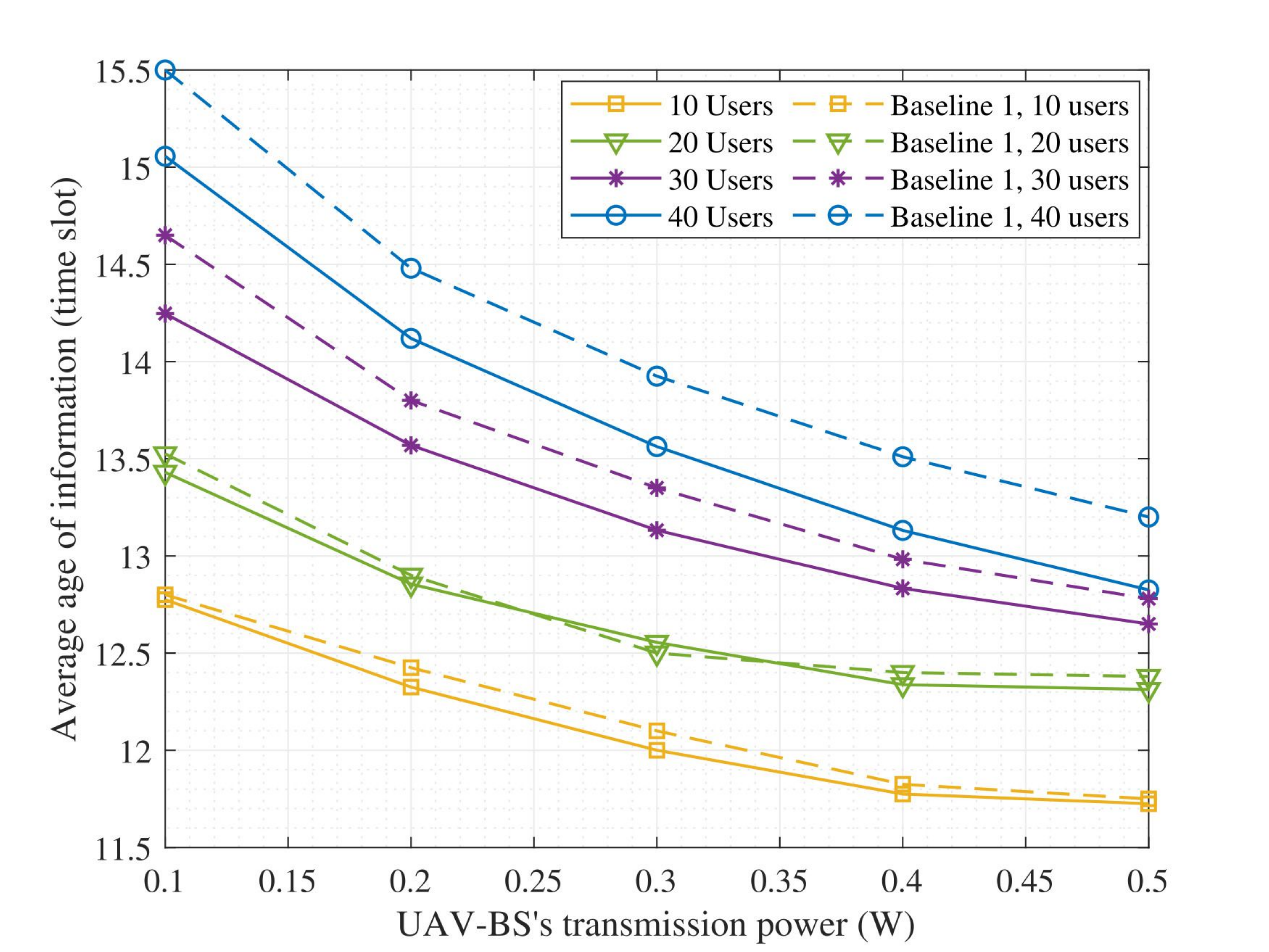}
		\caption{Impact of UAV-BSs' transmission power.}
		\label{fig:uavp}
	\end{subfigure}
	\caption{Impact of transmission power on network performance.}\label{fig:power}
\end{figure}

In this scenario, we analyze the impact of transmission power together with number of users on the average AoI and compare it with a baseline method, referered to as Baseline 1. In Baseline 1, tasks are sent without allocating transmission power but with their maximum power. As we can see in Fig. \ref{fig:userp}, the AoI-based performance of NTN network improves by $9.98 \%$, $9.92 \%$, $10.86 \%$, and $12.48 \%$, for $10$, $20$, $30$, and $40$ users, respectively, and $10.81 \%$ in overall, in the case of increasing user's transmission power from $0.1$ $\text{W}$ to $0.2$ $ \text{W}$. Comparing with Baseline 1, for the mentioned number of users, it is shown that allocating power to users achieves a lower AoI. Moreover, it can be seen that with higher number of users, the difference between two modes increases which is due to the increase of interference in Baseline 1, which results in lower data rate and higher AoI, accordingly. For the UAV-BSs’ transmission power, in Fig. \ref{fig:uavp}, the performance of the NTN network for $10$, $20$, $30$, and $40$ users improves by $8.22 \%$, $8.32 \%$, $11.20 \%$, and $14.82 \%$, respectively, and $10.64 \%$, in overall, by allocating larger transmission power. Similar to the case of user's transmission power, allocation of power to UAV-BSs gets better results than Baseline 1, in which all UAV-BSs send users' tasks by their maximum power in each time slot for the mentioned number of users. In particular, we can observe that the average AoI decreases by increasing the transmission power of both users and UAV-BSs. Such trends are in accordance with formulas (\ref{userdatarate}) and (\ref{rateU2U}).
\subsubsection{Task size}
Another interesting scenario is changing the size of tasks. In order to see this scenario better, we compare it with another baseline method, considered by previous works \cite{bhat2020throughput, moltafet2019power}, referred to as Baseline 2. In this case, it is assumed that the tasks are not separable and must be sent in one time slot and the computing capacity for the whole task is allocated either in UAV-BS or HAP but not partially at both. In other words, if the bandwidth and data rate assigned to the user is not sufficient to send a task, the task is not sent and the AoI increases, and also if the task reaches UAV-BS, but there is not enough capacity to process the entire task, AoI increases. Similarly, for sending data from UAV-BS to HAP, the same condition of sending and processing applies. Therefore, for this particular baseline scheme, we consider the following 
constraints:
\begin{align}
	&\alpha_{m, s}(t), \beta_{m,s}^{u}(t) = \{0, \alpha_{m, s}\},\\\nonumber
	&\eta_{m,s}(t), \theta_{m, s}^{u} (t)= \{0, 1\}.
\end{align}

\begin{figure}
	\centering
	\begin{minipage}{.5\textwidth}
		\centering
		\captionsetup{justification=centering}
		\includegraphics[width=\linewidth]{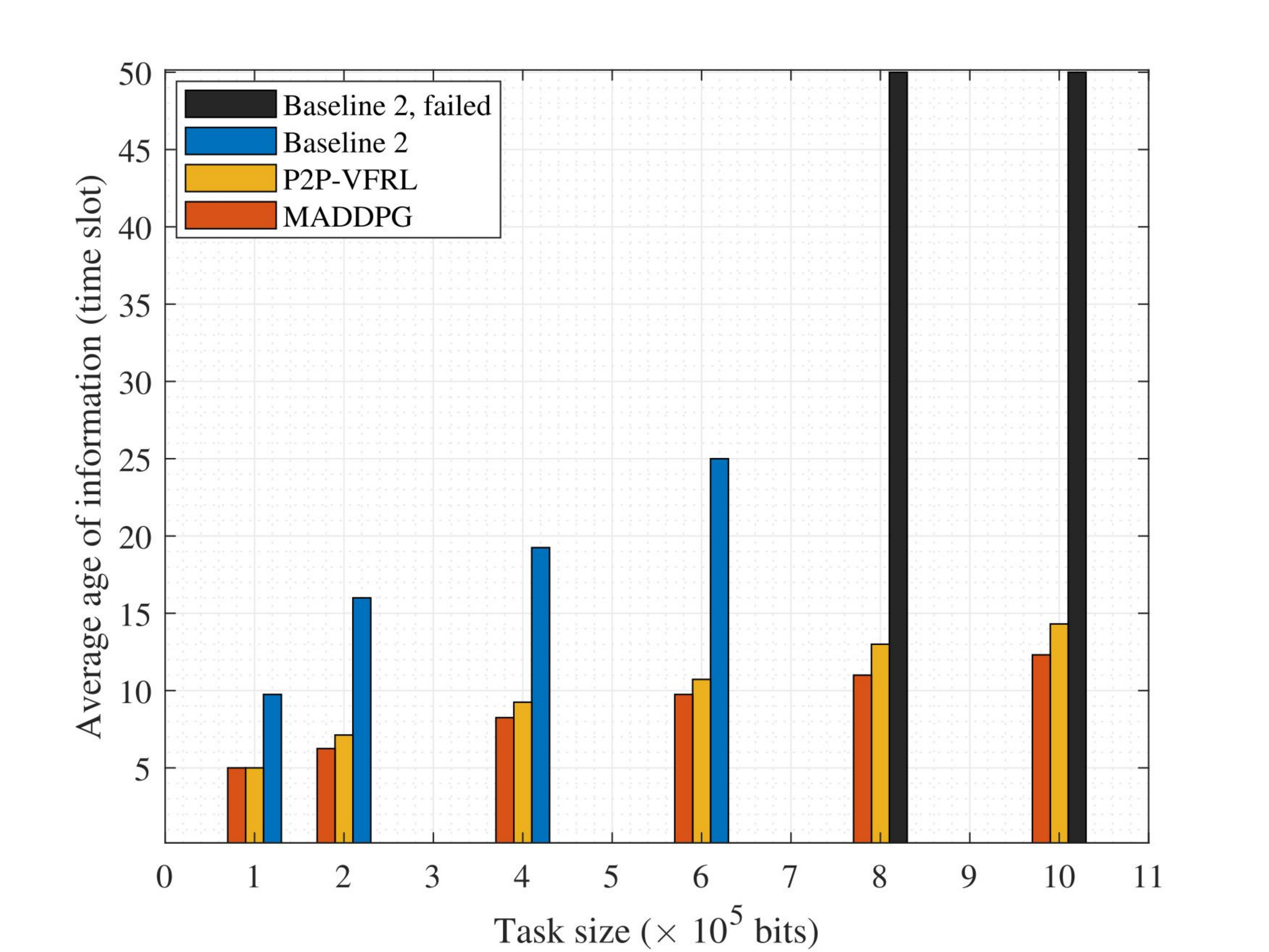}
		\captionof{figure}{\small{Comparing our two methods to Baseline 2 for different task sizes.}}
		\label{tasksize}
	\end{minipage}%
\end{figure}

This comparison is represented in Fig. \ref{tasksize}. As expected, with the increase in the size of the task, the value of AoI also increases, and this effect is more profound for Baseline 2 scheme. This is such that
for task sizes of $8\times 10^{5}$ and $10^{6}$ bits, the NTN network is not able to transmit at all which leads to AoI to become unacceptably large. This is referred to as transmission fail.
\subsubsection{Number of Subchannels}
\begin{figure}
	\begin{minipage}{.5\textwidth}
		\centering
		\captionsetup{justification=centering}
		\includegraphics[width=\linewidth]{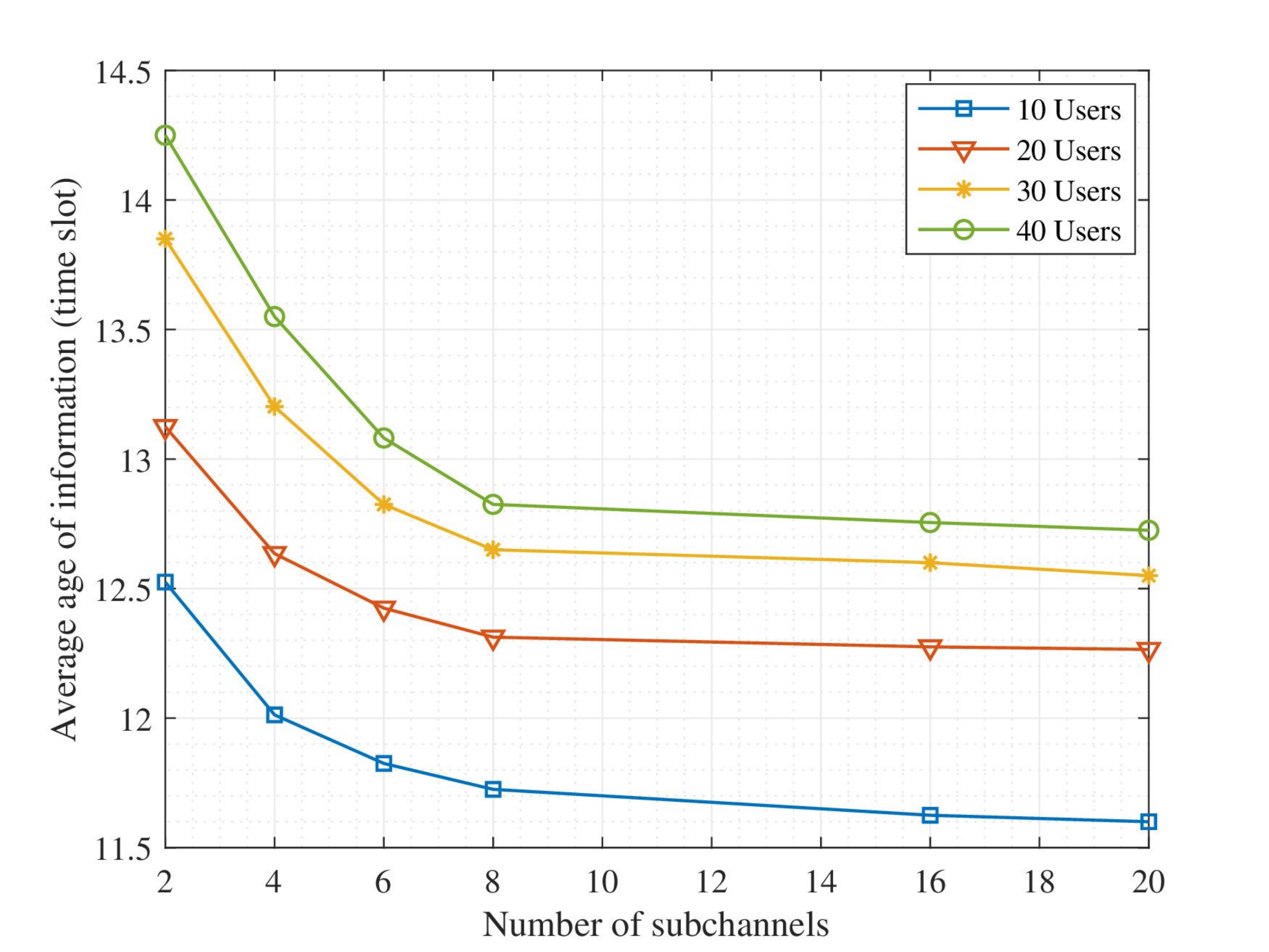}
		\captionof{figure}{\small{Impact of number of subchannels on AoI-based NTN performance.}}
		\label{subchannelimpact}
	\end{minipage}
\end{figure}

Fig. \ref{subchannelimpact} demonstrates the impact of the number of subchannels on the value of average AoI for different numbers of users. This figure shows that by increasing the number of subcarriers, and in other words, by increasing the bandwidth of each UAV-BS, more users send their tasks in each time slot, and as a result, more tasks are sent to the UAV-BS. The AoI-based performance for $10$, $20$, $30$, and $40$ users improved by $6.98 \%$, $7.01 \%$, $10.35\%$, and $11.98\%$, respectively. Nonetheless, as UAV-BSs are limited in their computation capacity, this parameter has a less profound effect on average AoI than the computing capacity of UAV-BS and HAP (Fig. \ref{fig:CPU}), as the average AoI does not improve that much for a larger number of subchannels.
\subsubsection{CSI Uncertainty}
To have a comparison of the effect of CSI imperfectness, we examine the average AoI by increasing the CSI uncertainty bound value in a range of $0\%$ to $20\%$. Channel gain has a direct impact on the data rate formula (\ref{userdatarate}). \textcolor{black}{As the CSI uncertainty bound increases, the amount of data rate allocated to each user decreases at each time slot which leads to a decrease in the portion of the task that is sent. As in our system model, the AoI is inversely proportional to data rate, the average AoI increases, as depicted in Figs. \ref{reward} and \ref{uncertaintyimpact}.}
\subsubsection{CSI Uncertainty}
\begin{figure}
	\centering
	\begin{minipage}{.5\textwidth}
		\centering
		\captionsetup{justification=centering}
		\includegraphics[width=\linewidth]{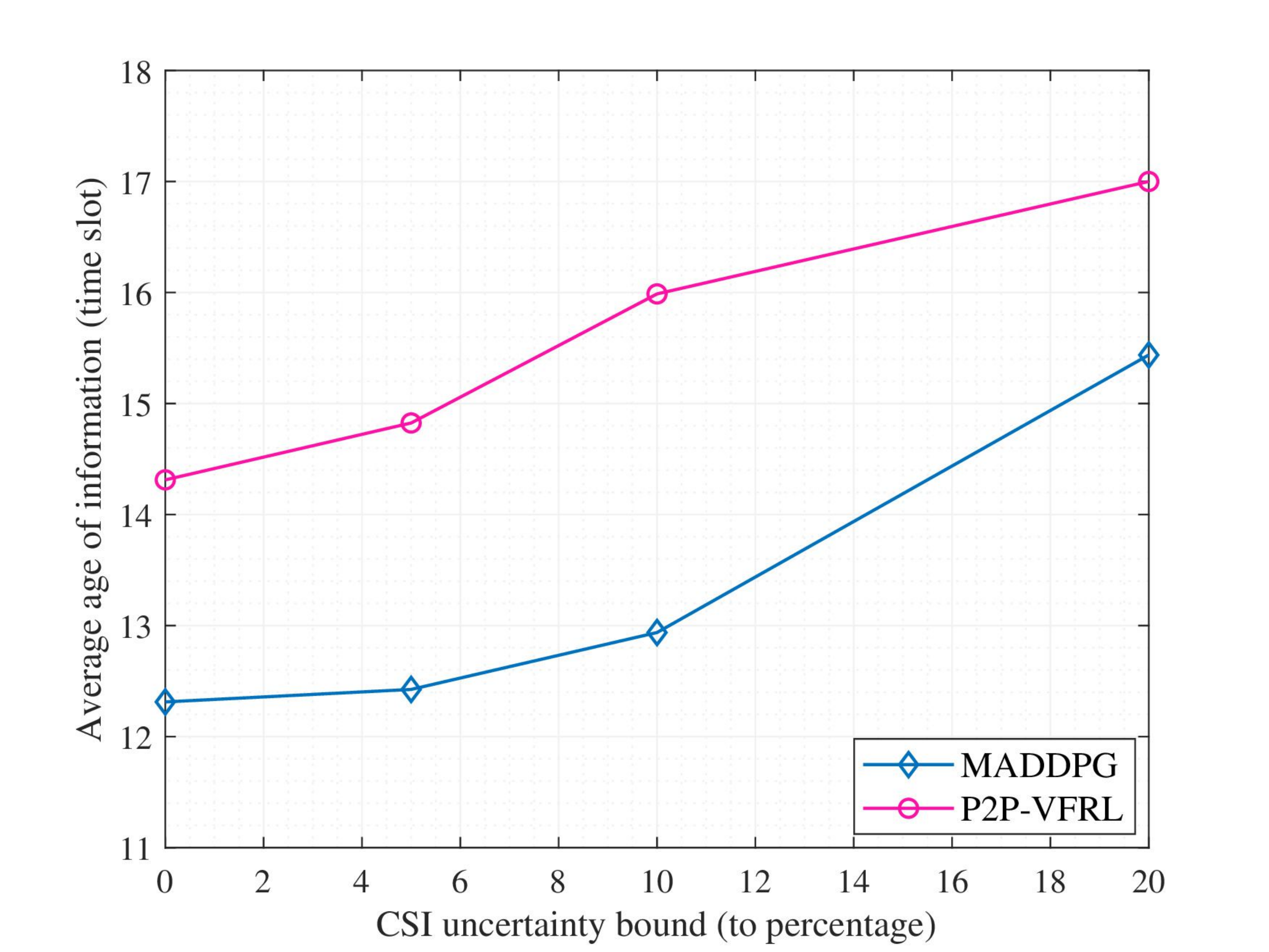}
		\captionof{figure}{\small{Impact of uncertainty on network performance.}}
		\label{uncertaintyimpact}
	\end{minipage}%
\end{figure}

To have a comparison of the effect of CSI imperfectness, we examine the average AoI by increasing the CSI uncertainty bound value in a range of $0\%$ to $20\%$. Channel gain has a direct impact on the data rate formula (\ref{userdatarate}). \textcolor{black}{As the CSI uncertainty bound increases, the amount of data rate allocated to each user decreases at each time slot which leads to a decrease in the portion of the task that is sent. As in our system model, the AoI is inversely proportional to data rate, the average AoI increases, as depicted in Figs. \ref{reward} and \ref{uncertaintyimpact}.}
\subsubsection{Trajectory}
\begin{figure}
	\begin{minipage}{.5\textwidth}
		\centering
		\captionsetup{justification=centering}
		\includegraphics[width=\linewidth]{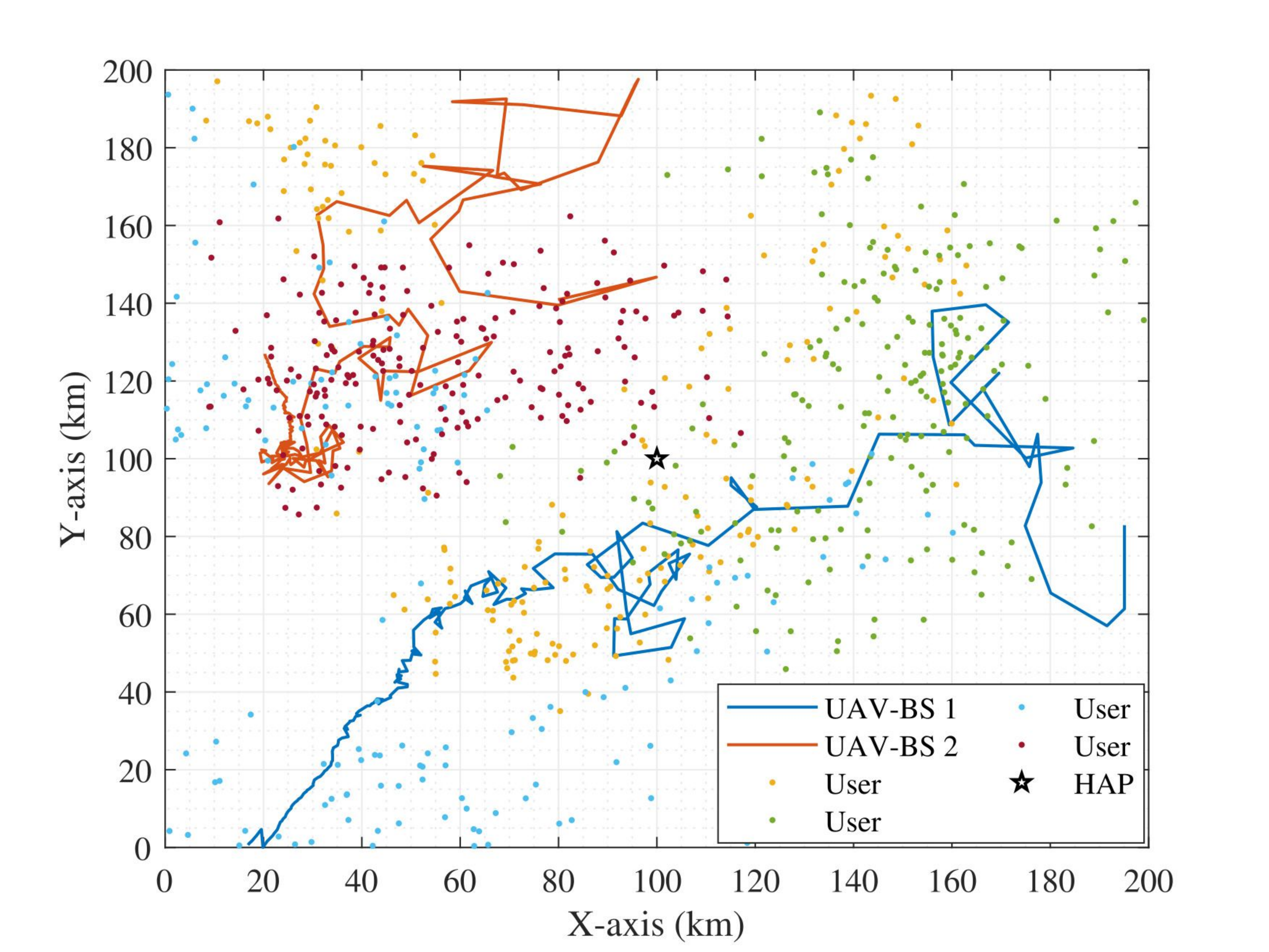}
		\captionof{figure}{\small{Trajectory of UAV-BSs and users' movement.}}
		\label{trajectory}
\end{minipage}
\end{figure}

To cover the mobile users in the area,  UAV-BSs are also mobile. In Fig. \ref{trajectory}, we considered an example which includes 2 UAV-BSs and 20 users and plot the corresponding trajectory for the UAV-BSs as well as the location of 4 of the users.  We assume that the location of the HAP is fixed during operation time. As can be seen, the UAB-BSs move so as to provide proper coverage. 
\section{Conclusion} \label{VI}
In this paper, we investigated the AoI sensitive two-tier aerial network including multiple UAV-BSs and a HAP, and provided online learning-based algorithms for channel, power, and computation capacity allocation and trajectory planning under CSI uncertainty in user-UAV-BS uplinks. The goal was to minimize the AoI of all users, who send their tasks to be processed. By adopting FL, and considering two types of agents in our environment, we developed a peer-to-peer online VFRL and compared it with well-known algorithm of MADDPG. We implemented a simulation setup with computationally-intensive applications. According to the results, for smaller task sizes, task scheduling reduces the average AoI approximately in half. For larger task size, the AoI gets unacceptably large if we rely on existing methods while by deploying task scheduling, the AoI decreases significantly and to an acceptable level. On the other hand, power allocation for both users and UAV-BSs has a marginal effect on the average AoI compared to using full transmission power. As a future work, one can consider a case where the UAV-BSs' selection ability is extended. For example, they can decide whether the remaining tasks in a UAV-BS have to be processed or sent to the HAP in the next time slot. In addition, choosing the optimized weights of the actor network parameters at each training step of P2P-VFRL algorithm can be considered as an action which may bring better results to reduce the performance gap between MADDPG and P2P-VFRL approaches.

%\appendices
%\section{Proof of the ...}

%\section{}

%\section*{Acknowledgment}


\ifCLASSOPTIONcaptionsoff
  \newpage
\fi

%\begin{thebibliography}{1}
% You can use other form of bib file by changing here... 


%\end{thebibliography}

%\begin{IEEEbiography}{Yuguang ``Michael'' Fang}
%\end{IEEEbiography}
\bibliographystyle{ieeetr}
\bibliography{IEEEabrv,Bibliography}

%It is not necessary to upload the biography when you submit your manuscript.

\end{document}